\documentclass[p5]{elsarticle}
\usepackage{amsmath}
\usepackage{amsfonts}
\usepackage{amsbsy}
\allowdisplaybreaks

\usepackage{graphics}
\usepackage{pstricks}
\usepackage{pst-coil}
\usepackage{pst-node}
\usepackage{pst-plot}
\usepackage{amssymb}

\DeclareMathOperator{\sgn}{sgn}
\DeclareMathOperator{\He}{H}
\DeclareMathOperator{\D}{D}

\renewcommand{\Re}{\operatorname{Re}}
\renewcommand{\Im}{\operatorname{Im}}

\newcommand{\eq}[1]{\eqref{#1}}
\newcommand{\1}{\ensuremath{\mathbf{1}}}
\newcommand{\0}{\ensuremath{\mathbf{0}}}
\renewcommand{\a}{\ensuremath{\mathbf{a}}}
\renewcommand{\b}{\ensuremath{\mathbf{b}}}

\renewcommand{\v}{\ensuremath{\mathbf{v}}}
\newcommand{\V}{\ensuremath{\mathbf{V}}}
\renewcommand{\d}{\ensuremath{\partial}}
\renewcommand{\r}{\ensuremath{\mathbf{r}}}

\newcommand{\B}{\ensuremath{\mathbf{B}}}
\renewcommand{\c}{\ensuremath{\mathbf{c}}}

\newcommand{\F}{\ensuremath{\mathbf{F}}}

\newcommand{\Rm}{\ensuremath{ R_{\text{m}}}}
\newcommand{\vm}{\ensuremath{ \v_{\text{m}}}}
\newcommand{\Vm}{\ensuremath{ \V_{\text{m}}}}

\newcommand{\grad}{\ensuremath{\nabla}}
\newcommand{\f}{\ensuremath{\mathbf{f}}}

\newcommand{\curl}{\ensuremath{\nabla\times}}

\begin{document}

\title{Spherical
single-roll dynamos at large magnetic Reynolds numbers}
\author[cam1,cam2]{Henrik Latter}
\author[cam2]{David Ivers}
\address[cam1]{LERMA-LRA, D\'epartement de Physique, Ecole Normale Sup\'erieure,
24 rue Lhomond, Paris 75005, France}
\address[cam2]{School of Mathematics and Statistics, University of Sydney, NSW 2006,
Australia}

\begin{frontmatter}

\begin{abstract}
This paper concerns kinematic helical dynamos in a
spherical fluid body surrounded by an insulator. In particular, we
examine their behaviour in the regime of large magnetic Reynolds number $\Rm$, for which dynamo
action is usually concentrated upon a simple resonant stream-surface. The
dynamo eigensolutions are computed numerically for two representative
single-roll flows using a compact spherical harmonic decomposition and
fourth-order finite-differences in radius. These solutions are then compared
with the growth rates and eigenfunctions of the Gilbert and Ponty (2000)
large $\Rm$ asymptotic theory.
We find good agreement between the growth rates when  $\Rm\!>\!10^4$,
and
between the eigenfunctions when $\Rm\!>\!10^5$.

\end{abstract}

\end{frontmatter}

\section{Introduction}

We consider a class of kinematic dynamos in which the magnetic field
$\B$ is
generated by the steady helical motion $\v$ of an incompressible,
electrically-conducting fluid. Helical flows constitute some of the simplest and most
efficient mechanisms for the excitation of a seed magnetic field, as witnessed
in numerical simulation and exploited by laboratory experiments (Gailitis et
al.~1987, Dudley \& James 1989, Forest et al.~2002, Moss 2008, for example). Such flows are
also widespread in astrophysical fluid bodies, such as jets, the
convection zones of stars and, possibly, liquid planetary cores, where they
might appear as a slow meridional
circulation (Giles et al.~1997, Gough and McIntyre 1998, Olson et al.~1999,
Haber et al.~2002,
 Hartigan et al.~2005).

By far the simplest helical flow is the Ponomarenko dynamo (Ponamerenko 1973), which consists of
a solid
electrically-conducting cylinder of finite radius rotating with constant angular
velocity
$\Omega$ and translating with a constant axial velocity $V$, while its 
rigid (electrically-conducting)
exterior remains at rest.
The dynamo loop in this case comprises (a) generation of azimuthal
magnetic field
from radial magnetic field by the discontinuity in the rotation (the `omega effect'), and (b)
reciprocal generation of radial field from
azimuthal field by magnetic diffusion. Note that the additional longitudinal shearing
component $V$ is crucial, as it draws apart oppositely directed field
lines and
prevents flux expulsion. Ponomarenko (1973) determined the magnetic field in this case
analytically, showing that the field is concentrated at the velocity discontinuity
on the
cylinder boundary.
Helical dynamos with non-uniform $V$ and
$\Omega$ and arbitrary cross-section, such as
in a conducting fluid, are naturally more complicated
(but see the exact steady solutions of Lortz 1968, and Chen \& Milovich 1984).
 Nevertheless, they share the same
dynamo ingredients: differential rotation and magnetic diffusion. These ingredients
are especially conspicuous at very large magnetic Reynolds
number $\Rm$, a regime which can be probed successfully with asymptotic theory 
(Ruzmaikin et al.~1988, Gilbert 1988, Gilbert \& Ponty 2000, hereafter GP). In such models
 the importance of diffusion, in particular, is evident from the localisation of
potentially growing dynamo modes upon their critical surfaces, where the modes are
nearly convected by the flow. It is upon these surfaces that the modes exhibit the 
small-scale variation necessary for magnetic diffusion to work effectively
and replenish radial magnetic field. Not every stream-surface can
support a mode, but a `resonance
condition' selects the critical surface (or surfaces) upon which dynamo action
can occur. 

 The asymptotics of helical dynamos in an infinite
 cylinder have
 been generalised to helical dynamos in a sphere (GP). 
A spherical single-roll dynamo can be portrayed
as a
cylindrical helical flow with its two ends joined,
 bent into a donut, and deformed to fill a spherical
`container'. The theory predicts that at large $\Rm$ the dominant mechanism of field
 generation is of Ponamarenko type and, indeed, that the same asymptotic
 scalings hold.
 However, full numerical studies of the dynamo problem
 in spheres have mostly tested small or intermediate
  $\Rm$, primarily to establish the critical $\Rm$
 necessary for dynamo action (Dudley \& James 1989, Forest et
 al.~2002, Moss 2008). The very large $\Rm$ regime has not received the same
 attention, partly because it remains a challenging numerical task. Consequently,
 the GP asymptotic theory has not yet been numerically confirmed. This is the main
 project of our paper

We undertake an analysis of single-roll helical dynamos in a
sphere, surrounded by an insulator, at very large magnetic Reynolds numbers. This is accomplished
numerically by the solution of the magnetic induction equation with suitable
boundary conditions for two representative single-roll flows. 
 The method of solution is presented in Ivers
and Phillips (2003, 2008) and consists of approximating the problem as a
large-scale algebraic eigenvalue equation.
 The growth rates and
eigenfunctions so obtained are subsequently compared with the predictions of the
GP asymptotic theory. We find excellent agreement in the regime $\Rm>10^4$ for
the growth rates and angular frequency, and good agreement for the magnetic
field structure when $\Rm>10^5$. In particular, the magnetic field structure
clearly localises upon a specific stream-surface, in contrast to intermediate
$\Rm$ where the field is somewhat disordered. This confirms that spherical
single-roll dynamos are indeed of the Ponamerenko type: no other growing modes
were discovered. This is a point that can
deepen our understanding of spherical dynamos more 
generally and aid the analysis of more complex
spherical flows. In particular, it is an entry point into the study of
multiple rolls, whose magnetic field generation will also be influenced by the
Gailitis dynamo mechanism (Gailitis 1970, 1993, 1995, Moss 2006).

The layout of the paper is as follows. In Section 2, the formal dynamo problem
is stated --- its governing equations, parameters,
and boundary conditions --- and the two flows we examine are presented. A brief
summary of the GP asymptotic results at large $\Rm$ follows in Section 3, while
their derivation is given in Appendix A, in the Supplemental Material,
 alongside a method to obtain higher order
terms.
Our results
and a comparison of the two approaches are given in Section 4, and conclusions
drawn in Section 5.

\section{Governing equations and setup}

\subsection{Problem formulation}
Consider a sphere of conducting fluid $V$ with radius $a$ and uniform magnetic
diffusivity $\eta$ surrounded by an
insulator $V^c$. Suppose that the fluid is undergoing time-steady
incompressible
 motions according to the velocity $\v$. Consequently, the magnetic field in the conducting fluid is governed by the
non-dimensionalised
 induction equation,
\begin{equation}\label{1.1}
  \d_\tau\B = \nabla ^2\B + \Rm \nabla \times (\v \times \B)
\end{equation}
where the magnetic Reynolds number $\Rm=\mathcal{V}a/\eta$ is defined in
terms of a
typical velocity $\mathcal{V}$, the radius $a$, and
$\eta$. The time $\tau$ is scaled on the magnetic diffusion time
$\mathcal{L}^2/\eta$ and space by $a$. The
magnetic field $\B$ is solenoidal everywhere,
\begin{equation}\label{1.2}
    \nabla\cdot\B = 0\,.
\end{equation}
Because the flow is steady, the
magnetic field can be expressed as a linear superposition of time-separable
solutions of the
form
\begin{equation}\label{1.2.2}
    \B(\r,\tau) = \B(\r)e^{\lambda\tau}\,,
\end{equation}
possibly with polynomial factors of time in degenerate cases. In addition, we
must supply suitable boundary conditions 
\begin{equation}\label{1.3}
  [\B]_\Sigma = \0\,,\qquad
  \curl\B = \0 \quad \text{in $V^c$,}\qquad
  \B \to \0 \quad \text{as $r\to\infty$,}
  \end{equation}
where $\Sigma$ is the surface of the sphere. This leads to an
eigenvalue
problem for the (complex) growth rate $\lambda$ and the associated eigenfunction.
For a given
flow $\v$, the growth rate $\lambda$ is a function of the magnetic Reynolds number
$\Rm$. When $\Re\lambda>0$, the
flow acts as a kinematic dynamo, i.e. the non-magnetic state $\B=\0$ is unstable to magnetic
perturbations.

\subsection{Representation of the helical flow}
We use spherical coordinates whereby the radius, polar angle, and azimuthal
angle are denoted by $r$, $\theta$, and $\phi$, with their accompanying unit
vectors given by $\1_r$,
$\1_\theta$, and $\1_\phi$, respectively.
An axisymmetric single-roll helical flows may be represented by
\begin{equation} \label{1.7}
\v= \sigma\,\Vm + W(r,\theta)\,r\,\sin\theta\,\1_\phi, 
\end{equation}
where $\Vm$ is the (scaled) meridional velocity, $W$ is the azimuthal \emph{angular}
speed, and $\sigma$ is a parameter measuring the relative strengths of the
meridional and azimuthal motion.
The
meridional flow $\Vm$ can be written in terms of a stream function $\Psi$ by
\begin{equation} \label{1.8}
  \Vm = -\frac{\d_\theta\Psi}{r^2\sin\theta}\1_r +
\frac{\d_r\Psi}{r\sin\theta}\1_\theta
  = \nabla\phi\times\nabla\Psi
  = -\nabla\times\frac{\Psi}{r\sin\theta}\1_\phi \,.
\end{equation}
The streamlines of $\Vm$ in a meridional plane are the level contours of $\Psi$ and
circle a
local minimum (maximum) of $\Psi$ in the clockwise (counter-clockwise)
direction. We also introduce the `unscaled' meridional velocity $\vm=\sigma\Vm$ and
stream function $\psi=\sigma\Psi$, which are more convenient when describing the
asymptotic theory.

\begin{figure}
\begin{center}
\scalebox{0.7}{\includegraphics{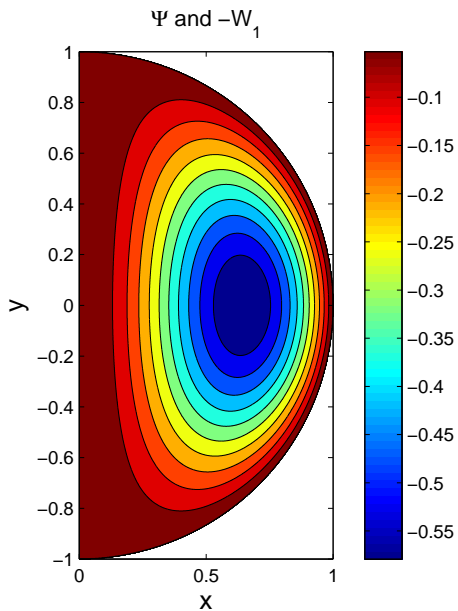}}
\end{center}
\caption{A representation of the meridional circulation shared by both flow 1
  and 2. The coloured isocontours of the meridional stream function $\Psi$ are
  plotted. The negative of this function is also the azimuthal angular
  speed of flow 1, as $W_1=-\Psi$.}
\end{figure}

In this paper, we examine two representative single-roll flows, $\v_1$ and
$\v_2$. The two flows share the same meridional velocity $\Vm$ but differ in
the azimuthal component $W\,r\,\sin\theta$. We set the stream function $\Psi$ according to
\begin{equation} \label{1.10}
\Psi(r,\theta) = - r \sin\pi r \sin^2\theta\,,
\end{equation}
but set
\begin{equation}\label{1.9}
    W_1 =  r \sin\pi r\,\sin^2\theta\,,\qquad
    W_2 = \frac{\sin\pi r}{r}\,.
\end{equation}
Flow 1 therefore possesses the restricted form $W_1=W_1(\Psi)=-\Psi$, while
flow 2
has a more general form.  The
restricted form of the local angular velocity simplifies the asymptotic theory
substantially.

\begin{figure}
\begin{center}
\scalebox{0.7}{\includegraphics{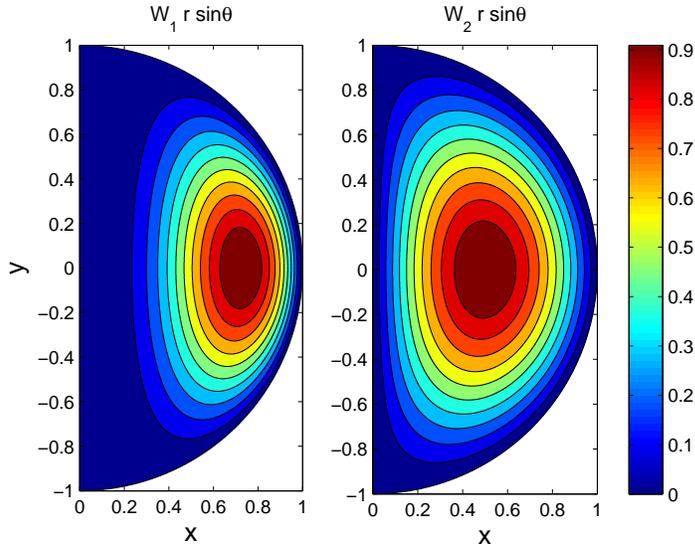}}
\end{center}
\caption{A representation of the two azimuthal flows considered. The left panel
  shows the azimuthal velocity of flow 1, $W_1\,r\sin\theta$, while the right panel
  shows that of flow 2, $W_2\,r\sin\theta$.}
\end{figure}

In Figs 1 and 2 the two flows are represented. Figure 1 shows the isocontours
of $\Psi$, the meridional stream function common to both flows.
 As the figure indicates, fluid
circulates in a clockwise direction around a central stagnation point located at $r\approx 0.646$. In
the case of flow 1, for which $W_1=-\Psi$, the isocontours of the azimuthal rotation
coincide with the isocontours of the meridional rotation. Therefore a fluid element upon a
given streamsurface will not only circulate at a constant speed in the meridional
plane it will also travel at a constant angular 
speed in the azimuthal direction. It is then easier
 to find flow trajectories that form closed loops.

 This
is not the case for the more complicated flow 2, whose azimuthal 
angular speed exhibits a purely radial
profile. Thus surfaces of constant rotation are spherical shells.
This means a fluid element in
flow 2 will experience different azimuthal angular velocities as it traverses
a meridional streamline. The actual azimuthal velocities, $W r\sin\theta$, of both flows are
plotted in Fig.~2.

The full numerical method we employ requires that the velocities are split
into toroidal and poloidal components and expanded in spherical
harmonics:
$$ \v= \sum_{m,n} \mathbf{t}_n^m \,+\,\sum_{m,n}\mathbf{s}_n^m\,, $$
where the toroidal components are given by
$\mathbf{t}_n^m = \curl (t_n^mY_n^m\r)$ and the poloidal components by $\mathbf{s}_n^m = \curl\curl
(s_n^mY_n^m\r)$,
where $Y_n^m$ is a spherical harmonic (see Ivers \& Phillips 2003, 2008).

 Flow 1 has
 the
poloidal-toroidal spectral decomposition $\v_1 = \sigma\, \mathbf{s}_1^0 + \mathbf{t}_1^0 +
\mathbf{t}_3^0$, in which the radial functions are
\begin{equation} \label{1.11}
s_1^0 = \frac{\sin\pi r}{\sqrt{3}}\,,\qquad
t_1^0 = -\frac{4}{5\sqrt{3}}\, r^2 \sin\pi r\,,\qquad
t_3^0 = \frac{2}{15\sqrt{7}}\, r^2 \sin\pi r\,,
\end{equation}
and the spherical harmonics are $Y_1^0=\sqrt{3}\,\cos\theta$ and
$Y_3^0=\tfrac{1}{2}\sqrt{7}\,\cos\theta(5\cos^2\theta-3)$.

The second flow $\v_2$ is the single-roll flow of Dudley \& James (1989).
 This flow has the poloidal-toroidal spectral
form $\v_2= \sigma\, \mathbf{s}_1^0 + \mathbf{t}_1^0$, where
\begin{equation} \label{1.12}
s_1^0 = \frac{\sin\pi r}{\sqrt{3}}\,,\qquad t_1^0 = \sin\pi r\,.
\end{equation}
 Although $\v_1$ has the simpler restricted form of the local angular velocity, its
spherical harmonic representation is actually more complicated than $\v_2$.

\section{The GP asymptotic theory}

This section summarises the main features and results of the GP asymptotic
theory, undertaken at large $\Rm$.
Here we give, without proof, the
leading order terms for the growth rates at $\mathcal{O}(\Rm^{-1/2})$ and the leading order contribution to
the magnetic eigenfunction. The derivations of
these expressions are placed in the Appendix A of the Supplemental Material 
for reference. There we also present a method whereby higher order terms in the
asymptotic expansion can be calculated.

\subsection{Toroidal co-ordinates for axisymmetric helical flows}

The structure of the helical flow \eq{1.7}, i.e.\ the topology of its streamlines and its
differential
rotation, can be exploited by a toroidal coordinate system which simplifies the
advection
operator. The first two coordinates are determined by the meridional flow. The
stream function
$\psi$ is one coordinate (note that we use the unscaled stream function).
 The second is an angle coordinate $\vartheta$ defined as
follows. If
$T$ is the period for a fluid particle to traverse once the closed streamline
$\psi=\psi_o$, then $T=T(\psi)$ and
\begin{align} \label{2.3}
  d\vartheta \equiv \frac{2\pi}{T} dt = \Omega\frac{d\ell}{q} = \Omega\frac{\vm\cdot
d\r}{q^2}
  = \Omega\frac{r d\theta}{v_\theta} = \Omega\frac{dr}{v_r}\,,
\end{align}
where $\Omega=2\pi/T$ is the angular frequency, $d\ell$ is the element of arc-length
travelled
in a time $dt$ and $q = |\vm|$. Thus
$d\vartheta/dt=\Omega$ is constant on streamlines. For the two flows we consider we fix
$\vartheta=0$ on the $s$-axis $(\theta = \pi/2)$, since the stagnation points of their
meridional parts occur there. Clearly $\vartheta$ changes by $2\pi$ in one full
traversal of
the closed streamline. We denote by an overbar, or pair of angle brackets, the average
around the
streamline $\psi=\psi_o$, defined by
\[ \overline{f} = \langle f \rangle \equiv \frac{1}{2\pi} \oint_{\psi=\psi_o} f
d\vartheta\,, \]
where $f$ is a function of the meridional coordinates $\psi$ and $\vartheta$ 
and the integration is around the
streamline $\psi=\psi_o$. We can thus take any quantity $F(\vartheta,\psi)$
and compute a mean
component
independent of $\vartheta$, i.e. $\overline{F}=\overline{F}(\psi)$, and 
a
`fluctuating' component, $\widetilde{F}(\psi,\vartheta) \equiv
F(\psi,\vartheta)-\overline{F}$.

Having specified these details, we replace the azimuthal angle $\phi$ by a third
coordinate
$\zeta$ defined by
\begin{equation}\label{2.4}
  \zeta(\psi,\vartheta,\phi) \equiv \phi - Z (\psi,\vartheta)\,,\qquad
  Z = \frac{1}{\Omega(\psi)} \int_0^\vartheta
\widetilde{W}(\psi,\vartheta^*)\,d\vartheta^*\,.
\end{equation}
The level surfaces of $\zeta$ are hence distorted azimuthal planes

The coordinate system $(\psi,\vartheta,\zeta)$ naturally gives rise to the two
right-handed
vector bases, $(\nabla\psi,\nabla\vartheta,\nabla\zeta)$ and
$(\f_\psi,\f_\vartheta,\f_\zeta)$,
where $\f_\psi = \d\r/\d\psi$, $\f_\vartheta =\d\r/\d\vartheta$,
$\f_\zeta=\d\r/\d\zeta$ and
$\r$ is the position vector. It is a useful shorthand to also denote the coordinates by
$\psi^i$ with indices $i=1,2,3$, and the two bases by $\nabla\psi^i$ and $\f_i$. The
bases are
reciprocal, $\f_i\cdot\nabla\psi^j=\delta_i^j$, and related by
\begin{equation}\label{2.6}
 \f_\psi = J\nabla\vartheta\times\nabla\zeta\,,\quad
 \f_\vartheta = J\nabla\zeta\times\nabla\psi\,,\quad
 \f_\zeta = J\nabla\psi\times\nabla\vartheta\,,
\end{equation}
where the Jacobian $J$ of the transformation to $(\psi,\vartheta,\zeta)$ is given by
\begin{equation} \label{2.7}
  J = \f_\psi\times\f_\vartheta\cdot\f_\zeta =
  (\nabla\psi\times\nabla\vartheta\cdot\nabla\zeta)^{-1} = \Omega^{-1}.
\end{equation}
Using the properties of the flux function, equations \eq{2.6} may be simplified to
\begin{equation} \label{2.8}
\f_\psi = \Omega^{-1} \nabla\vartheta\times\nabla\phi + \f_\zeta\d_\psi
Z\,,\quad
\f_\vartheta = \Omega^{-1} \nabla\phi\times\nabla\psi + \f_\zeta\d_\vartheta
Z\,,\quad
\f_\zeta = r\sin\theta\1_\phi\,.
\end{equation}
Finally, using the properties of reciprocal bases the velocity and the magnetic field may be
written as
\begin{equation} \label{2.9}
\v = \Omega(\psi) \f_\vartheta + \overline{W}(\psi) \f_\zeta\,,\qquad
\B = B_\psi \f_\psi + B_\vartheta \f_\vartheta + B_\zeta \f_\zeta\,.
\end{equation}
The advection operator is expressed as
\begin{equation} \label{2.5}
  D_t \equiv \d_t + \v\cdot\nabla = \d_t + \Omega(\psi) \d_\vartheta + \overline{W}(\psi) \d_\zeta\,.
\end{equation}
Its sole dependence on $\psi$ is essential in the asymptotic theory. 

The magnetic induction equation scaled on the turn-over timescale $t=\tau\Rm$ is
\begin{equation} \label{2.14}
\d_t\B = \nabla\times(\v\times\B) + \varepsilon^4\nabla^2\B\,,
\end{equation}
where $\varepsilon \equiv \Rm^{-1/4}$. The contravariant components of \eq{2.14}
with respect to the new coordinates are
\begin{align}
  D_t B_\psi &= \varepsilon^4 \nabla\psi\cdot\nabla^2\B
  \label{2.15}\\
  D_t B_\vartheta - \Omega'(\psi) B_\psi &=
\varepsilon^4\nabla\vartheta\cdot\nabla^2 \B
  \label{2.16}\\
  D_t B_\zeta - \overline{W}'(\psi) B_\psi &=
\varepsilon^4\nabla\zeta\cdot\nabla^2 \B\,.
  \label{2.17}
\end{align}
The primes indicate derivatives with respect $\psi$. As noted above, regeneration of
the
magnetic field component $B_\psi$ is solely due to diffusion, but regeneration of
$B_\vartheta$
and $B_\zeta$ is partly due to distortion of $B_\psi$ by meridional and azimuthal
differential rotation, respectively.

\subsection{Asymptotic scalings}

The GP theory is developed in the large $\Rm$ regime, i.e.\ as  $\varepsilon \to
0$. The leading order solution is decomposed into dynamo modes of the form
$B_\psi,B_\vartheta,B_\zeta\propto
e^{im\zeta + ik\vartheta+ (p+i\omega)t}$. Thus $p+i\omega = \lambda/\Rm$. The
constants $m$ and $k$ must
be
integers for solutions single-valued in $\zeta$ and $\vartheta$. The following scalings are
subsequently adopted
(Ruzmaikin et al.~1988)
\begin{equation} \label{2.18}
 \varepsilon^2 B_\psi \sim  B_\vartheta \sim B_\zeta\,,\quad
 p = {\mathcal O}(\varepsilon^2)\,,\quad m,k = {\mathcal O}(1)\,,
\end{equation}
moreover, it is assumed that modes localise upon a stream
surface
$\psi=\psi_o$ in a layer of thickness ${\mathcal O}(\varepsilon)$. Thus $\vartheta$-
and
$\zeta$-derivatives are ${\cal O}(1)$, but $\psi$-derivatives are ${\cal
O}(\varepsilon^{-1})$.
This suggests a new variable $\varUpsilon$ defined through
\begin{equation} \label{2.19}
  \psi = \psi_o + \varepsilon\varUpsilon\,.
\end{equation}
so that $\varUpsilon$-derivatives are ${\mathcal O}(1)$. 

The magnetic solution and growth rates are subsequently
 expanded in powers of $\epsilon$, and a particular streamline $\psi_0$ is
 chosen and equilibrium quantities depending on $\psi$ expanded in Taylor
 series about this streamline. These are substituted into the governing
 equations and we collect terms order by order solving each set of equations as we go. These
 details are summarised in Appendix A of the Supplemental Material.
 In brief, solvability of the zeroth order equations forces a
 given dynamo mode to localise on its
 critical stream surface, where it will be convected with the flow.
 Solvability of the first order equations, yields a `resonance condition'
 which selects which stream surface (or surfaces) can actually harbour such
 dynamo modes. Solvability of the second order equations provides the leading
 order $\varUpsilon$-structure of the modes and the leading order growth rate.

\subsection{The leading order asymptotic solution}

In the following expressions we transform from $\psi$ to $\Psi$ and thus make
explicit the dependence on the tuning parameter $\sigma$. 

\subsubsection{Eigenfunctions}

To dominant order
spatially, dynamo modes take the form
\begin{equation} \label{fuck1}
\B \,\propto\, \mathbf{a}\,  D_n(\varUpsilon/\kappa)\,
e^{ik\vartheta+im\zeta+i\omega t+pt} 
\end{equation}
where $D_n(z)$ is the parabolic cylinder function of order $n$ (Abramowitz \&
Stegun 1972), with 
\begin{equation}\label{fuck2}
\mathbf{a} = \sigma\Omega'_o\,\mathbf{f}_\vartheta +
\overline{W}'_o\,\mathbf{f}_\zeta = \left\{ \frac{\Omega'_o}{ \Omega_o}\,\Vm + \frac{1}{\sigma}\left( \frac{\Omega'_o}{\Omega_o}\,\widetilde{W}
+ \overline{W}'_o \right)\,r\sin\theta\,\1_\phi \right\},
\end{equation}
and
\[
1/\kappa = \sqrt[4]{|\Pi''_o|/\overline{\gamma}_0}\bigg(
\sqrt{\tfrac{\sqrt{2}+1}{2}}
+ i\sqrt{\tfrac{\sqrt{2}-1}{2}}\ \sgn\Pi''_o\bigg)\,,
\]
where $\Pi=\sigma k\Omega + m\overline{W}$, $\overline{\gamma}_0=\langle|\nabla\Psi_o|^2
\rangle$, and a prime now indicates differentiation with respect to $\Psi$. 
The parabolic cylinder functions impart a Gaussian-like structure about the
resonant curve,
with spatial oscillations of rapidly
diminishing amplitude as distance $\varUpsilon$ from the curve increases. Higher $n$
modes
display more complex spatially varying behaviour within the envelope of the stream
surface
localization.

\subsubsection{Growth rates}

The real part of the growth rate is
\begin{equation} \label{2.57}
  p = \pm \varepsilon^2 \sqrt{| k \overline{\mu}_b + m \overline{\mu}_c|
|\Omega_o'|}
    - (n+\tfrac{1}{2})\varepsilon^2\sqrt{| k\sigma\Omega_o'' + m\overline{W}_o''|
\overline{\gamma}_0}
    + \mathcal{O}(\varepsilon^{4})
\end{equation}
and the angular frequencies are
\begin{multline} \label{2.58}
  \omega = -\sigma k \Omega_o - m \overline{W}_o \pm \varepsilon^2
  \sqrt{| k \overline{\mu}_b + m \overline{\mu}_c| |\Omega_o'|}\,\sgn
  \left[ ( k \overline{\mu}_b + m \overline{\mu}_c)\Omega_o' \right]
  \\
  - (n+\tfrac{1}{2})\varepsilon^2\sqrt{|\sigma k\Omega_{o}'' + m\overline{W}_o''|
\overline{\gamma}_0}\,
  \sgn\left(\sigma k\Omega_o'' + m\overline{W}_o''\right) +
\mathcal{O}(\varepsilon^4)\,,
\end{multline}
where the quantities appearing are evaluated upon the resonant stream surface
$\Psi=\Psi_o$.
The $n=0$ mode is the fastest growing magnetic field mode to dominant order for any
$m$ and
$k$. The geometric $\mu$ terms are 
$$ \mu_b= \sigma\nabla\Psi\cdot(\nabla\vartheta\cdot\nabla\mathbf{f}_\vartheta)\,,
\qquad \mu_c= \nabla\Psi\cdot(\nabla\zeta\cdot\nabla\mathbf{\vartheta}).$$

In addition GP add terms of higher order $\varepsilon^4$ to the expression for
$p$ arguing that these are the most important  at this order. The additional terms are
 $$ -\varepsilon^4\left( k^2\overline{\beta}_k +
m^2\overline{\beta}_m +
2mk\overline{\beta}_{mk}\right)\,. $$ 
The $\beta$'s, which are independent of
$\sigma$,
are given by
\begin{align}
\beta_k &= |\nabla\vartheta|^2 -\frac{1}
{\overline{\gamma}_0}(\nabla\Psi\cdot\nabla\vartheta)^2
\label{2.59}\\
\beta_m &= |\nabla\zeta|^2 -\frac{1}
{\overline{\gamma}_0}(\nabla\Psi\cdot\nabla\zeta)^2
\label{2.60}\\
\beta_{mk} &= \nabla\vartheta\cdot\nabla\zeta -
\frac{1}{\overline{\gamma}_0}(\nabla\Psi\cdot\nabla\vartheta)(\nabla\Psi\cdot\nabla\zeta)\,,
\label{2.61}
\end{align}

\subsubsection{Resonance condition}

The expression for the mode frequency Eq.~\eqref{2.58}, shows
that dynamo modes are (to leading order)
 `advected' by the flow at the streamline upon
which they localise. That is to say, $\omega \approx -\B\cdot\nabla\v$. 
It follows that the resonant streamsurface is a magnetic critical layer,
and modes in its immeditae vicinity will naturally 
exhibit rapid spatial oscillations, though these are regularised by the (small) magnetic
diffusion upon the critical streamsurface itself. At large $\Rm$ these
oscillations are crucial to dynamo action, because they provide sufficiently
steep spatial gradients for the small resistivity to work efficiently, and
regenerate $B_\psi$. They hence close the dynamo loop begun by the
differential rotation across the layer.

We have not yet stated which streamline a given dynamo mode will prefer,
upon which the resonance condition holds. This condition may be written
as
\begin{equation} \label{2.62}
  k\sigma\Omega'_o + m\overline{W}'_o = 0\,,\qquad
  \sigma^2\Pi''_o = k\sigma\Omega''_o + m\overline{W}''_o \neq 0\,.
\end{equation}
In general, we find that $\Pi''_o<0$, which indicates that the resonant streamline
corresponds
to the maximal helical gradient of the magnetic mode.
 But Equation \eq{2.62}(a) is also the
condition for the closure of the magnetic field lines on the surface $\psi=\psi_o$
to leading
order, as the following argument shows.

 Since $d\mathbf{r} = \f_\psi d\psi + \f_\vartheta d\vartheta + \f_\zeta
d\zeta$, the
equation for the magnetic field lines, $\B\times d\mathbf{r} = \0$, reduces to
\[ \frac{d\psi}{B_\psi} = \frac{d\vartheta}{B_\vartheta} = \frac{d\zeta}{B_\zeta}\,. \]
For the magnetic field Eqs \eqref{fuck1}-\eqref{fuck2}, the field lines to leading order are
\[ \Psi=\Psi_o\,, \qquad \vartheta-\vartheta_o =
\frac{\overline{W}'_o}{\sigma\Omega'_o}(\zeta-\zeta_o)\,, \]
where $(\Psi_o,\vartheta_o,\zeta_o)$ is a given point on the field line. The
magnetic field
line is closed if there are integers $k$, $m$ such that $\vartheta-\vartheta_o=-2\pi
k$ and
$\zeta-\zeta_o=2\pi m$, which give the resonance condition \eq{2.62}(a). So,
unsurprisingly, a
resonant surface also corresponds to the spatial localisation for which a magnetic mode
reinforces itself. The resonance condition also ensures the $\varepsilon^0$-magnetic
field is
solenoidal.

\subsection{Discussion}

In practice, it is simplest to stipulate the ratio $(m/k)$ and the streamline
$\Psi$ and then compute the $\sigma$ necessary for this $\Psi$ to be resonant
from \eqref{2.62}. Therefore,
\begin{equation}\label{2.63}
    \sigma=-\left(\frac{m}{k}\right)\left(\frac{\overline{W}'_o}{\Omega'_o}\right)=\sigma(m/k,\Psi_o)\,,
\end{equation}
and $\sigma$ may be interpreted as an adjustable `tuning' parameter,
 permitting
 modes on any
 streamline we choose.

However, it is also instructive to examine how the dynamo modes,
and their resonant streamlines change, as $\sigma$ varies. The
parameter $\sigma$ controls the geometry of the helical flow by
establishing the ratio of the meridional circulation's speed against the
azimuthal rotation, and hence directly influences the dynamo action.

Consider flow 2. According to Eq.~\eqref{2.63}
 each choice of $\sigma$ and $m/k$ will give a
single resonant curve $\Psi_o$. Now fix the pitch $m/k$ of the dynamo modes
under consideration; this means that as we vary $\sigma$ we also vary the resonant
streamsurface. On the other hand, for flow 2, it can be shown that, as $\Psi$ varies between 0 and
 its minimum value, the quantity
 $\overline{W}'_o/\Omega'_o$ varies monotonically between two nearby
 constants,
 the smaller
 associated with the outermost streamline upon the spherical boundary,
 and the larger with the stagnation
 point. From \eqref{2.63}, it then follows that
 there exists only a (narrow) interval of $\sigma$ for which a resonance on any
 streamline is possible. Each choice of $m/k$ furnishes a different interval
 of $\sigma$ but none of these overlap. A subinterval of each may permit
 magnetic growth. Therefore as we vary $\sigma$, and consequently modify
 the flow geometry, we encounter discrete `windows', or `resonance intervals',
  of magnetic field
 generation. Note that dynamo action is not possible in every interval, in
 particular
 for very small and very large $\sigma$. These require large $m$ or
 $k$ which violates the scaling assumptions of the asymptotic theory,
 Eq.~\ref{2.18}. In any case, Proctor's modification of the toroidal
 anti-dynamo theorem (Proctor 2004) suggests that there can be no magnetic
 growth for $\sigma< \varepsilon^{1/2}$, i.e.\ for flows almost entirely
 azimuthal. In the other limit, $\sigma$ large, which corresponds to a flow
 dominated by the meridional component, things are less clear. For certain
 stream functions dynamo action appears possible in the complete absence of the
 azimuthal motion ($\sigma\to\infty$), though the
 relationship between the flow and the boundary is crucial (Moss 2008).
 
In contrast, the simpler flow 1 admits dynamo action for a very wide range of
$\sigma$; moreover, multiple $m/k$ modes may grow concurrently. In other words,
the $\sigma$ resonance intervals overlap substantially. Plainly, a simpler
flow, in which both the azimuthal and meridional motion are closely related,
is the more propitious for magnetic generation. This follows from the fact
that the fluid trajectories can more easily join the magnetic field lines they
convect into closed loops. In circumstances where the profiles of 
$\Psi$ and $W$ are dissimilar (flow 2), this is more difficult to do, and can only
occur when their relative magnitudes are tuned appropriately (by $\sigma$).

Considering how vital it is to for the flow to close magnetic field lines in
 the large $\Rm$ limit,
 it is natural to enquire into the (possibly deleterious) influence of small velocity 
fluctuations superimposed upon the mean helical motion. Recent work on the
 cylindrical Ponomarenko dynamo shows that magnetic growth persists when
 the amplitudes of the helical flow has a small time-dependent (fluctuating)
 part.
 Dynamo action even can occur when the meridional and azimuthal
 fluctuations are slightly different functions of time, forcing the resonant
 curve to also change with time (Peyrot et al.~2007,
 2008). Similar behaviour undoubtedly carries over to the spherical single
 roll dynamos we consider. However, when the small velocity fluctuation is not only a
 function of time, but of space as well, dynamo action will most likely suffer. 
In such a flow the fluid trajectories will not normally close.
 Thus, like flow 2, the
 magnetic field lines they transport will not normally close, and
 the
 resonance condition will be more difficult to satisfy. Dynamo activity may still be
 possible in the limit of small fluctuation amplitude, as then the fluid
 trajectories may not deviate beyond the magnetic localisation, but this
 probably can only be
 checked with numerical simulations.

\section{Results}

We present results for each of the flows $\v_1$ and $\v_2$, corresponding to a
representative configuration of the parameters for the same resonant stream surface. For both
$\v_1$ and $\v_2$ this resonant curve is $\Psi_o=\Psi(r_s,\pi/2)\approx -0.20287$
with $r_s=0.93$. The resonance is ensured for given $m$ and $k$ by setting the tuning
parameter $\sigma$ according to \eq{2.63}. The resonance conditions for $\v_1$ and
$\v_2$
may be expressed as $\sigma=\sigma_1(m,k,\Psi)$ and $\sigma=\sigma_2(m,k,\Psi)$
respectively. We find that 
\begin{align}
\sigma_1(1,1,\Psi_0) \approx 0.1373, \quad
\sigma_1(2,1,\Psi_0)\approx 0.2747, \quad \sigma_2(1,1,\Psi_0)\approx 0.2050,
\end{align}
 to 4
significant figures. Moreover, for the flows we examine there is no degeneracy in
$\sigma$, i.e. for a given $\sigma$ there is only one possible set of
$(m,k,\Psi)$, and hence only one resonant curve for a given flow.

The different times $t$, $\tau$ of the asymptotic and numerical results are related by
$t/\tau=\Rm$. Thus to compare the asymptotic and numerical results, the numerical
growth
rates are divided by $\Rm$, i.e. $p=\Re\lambda/\Rm$, $\omega=\Im\lambda/\Rm$. Moreover,
asymptotic and numerical modes must be correctly matched. There is no difficulty
with the
azimuthal wavenumber $m$, since it coincides in the asymptotic and numerical results
for
$\v_1$, and also for $\v_2$, when $\zeta$ is decomposed and the factor $e^{im\phi}$ is
extracted from the eigenfunction. We assumed that, once the resonant curve and the
wavenumbers $m$, $k$ are chosen, which sets the tuning parameter $\sigma(m,k,\Psi)$,
the
collection of modes determined by the numerical eigenproblem correspond to the various
asymptotic $n$ modes. The strongest growing exact (numerical) mode was identified with
$n=0$.

\subsection{Numerical methods}

The calculation of the asymptotic solution for a given flow and its rendering
in spherical coordinates is not straightforward. This called for various
analytical tricks and numerical
techniques, a full explanation of which we give in Appendix B in the
Supplemental Material. Most of the
effort lay in computing the $\mu$ and $\beta$ coefficients and also in
determining the $\Psi$ derivatives of $\Omega$ and $\overline{W}$ at
$\Psi=\Psi_o$. 
These quantities, in fact, can all be expressed as contour integrals of various
kinds over all or part of the streamline $\Psi_o$. When these integrals are closed
 they can be numerically approximated with excellent
accuracy. The chief numerical parameter here is the number of
$(r,\,\theta)$ points used to discretise the resonant streamline $\Psi_0$; this was denoted
by $K$ (see Appendix B in the Supplemental Material).

On the other hand, the full dynamo problem presents a considerable numerical
challenge, especially in the large $\Rm$ regime when the magnetic structure 
 exhibits small-scale localised variation. In this limit extreme resolution is required to
 properly capture the dynamo modes, which leads to the
 eigensolution of enormous, albeit banded, matrices. A full
 explanation of the techniques employed to approximate the magnetic
 induction operator with such matrices can be found in Ivers \& Phillips (2003,
 2008). We summarise the approach below. 

Our numerical method uses a hybrid version of the spectral
 equations of James (1974), which are in a similar form to the
 poloidal-toroidal spectral equations derived by Bullard \& Gellman (1954). 
The magnetic field and velocity field are expanded in vector spherical
harmonics when they appear in the advection term, but otherwise are decomposed
into
 toroidal-poloidal components and then expanded in scalar spherical
 harmonics. Doing so requires us to compute fewer coupling integrals. The
 spectral expansion is truncated at some large order $N$, so that
 for a given azimuthal mode number $m$ there are $2(N-m+1)$ spherical harmonic
 functions.

The radial dependence is discretised using fourth-order finite-differences
over a uniform grid. The number of radial points is denoted by
$J+1$. A centred-finite difference formula were used at interior points and
one-sided formulas at the boundaries.

Truncation in the number of harmonics and in radius converts the problem into
a set of linear equations for $(2J+1)(N-m+1)$ coefficients plus the growth
rate $\lambda$. We solve this algebraic eigenvalue problem by inverse
iteration and the implictly restarted Arnoldi method using ARPACK (Sorensen 1992).
 The Arnoldi method is
particularly helpful in identifying the mode of fastest growth at large $\Rm$
because the eigenvalues in this limit bunch together in the complex plane: the
ratio of the real part of the growth rate to the imaginary part is
$\mathcal{O}(\Rm^{-1/2})$ as indicated by the asymptotic theory (see also
Table 2). Consequently, inverse
iteration has difficulty in converging to the eigenvalue of largest real part without a
good estimate of the true eigenvalue.

\begin{table}
\begin{center}
\begin{footnotesize}
\begin{tabular}{c|ccc|ccc}
\hline
    & \multicolumn{3}{c|}{\rule{0pt}{10pt}$\Re\lambda$} &
\multicolumn{3}{c}{$\Im\lambda$}  \\
\hline
$J\backslash N$ & 20 & 30 & 40 & 20 & 30 & 40 \\
\hline
200 & 687.7 & 688.2 & 688.3 & 16390.6 & 16391.6 & 16392.8 \\
400 & 688.2 & 687.5 & 687.5 & 16390.6 & 16391.9 & 16392.1 \\
800 & 688.2 & 687.5 & 687.5 & 16390.6 & 16392.0 & 16392.3 \\
\hline
\end{tabular}
\end{footnotesize}
\end{center}
\caption{Convergence of the dominant mode's eigenvalue $\lambda$ with the
    numerical parameters $J$ and
$N$. The number of radial grid points is $J+1$, the number of spherical
harmonic functions is $2(N-m+1)$. The growth rates are for $\v_2$ at
$\Rm=10^5$ and $m=1$.}
\end{table}

At very large magnetic Reynolds number ($\Rm\sim 10^5$), a converged solution (with respect to
resolution) requires extremely large $J$ and $N$. The largest matrix computed was
$64,040\times 64,040$ for $J=800$,
$N=40$, and
$m=1$. Convergence of the eigenvalue $\lambda$ of largest real part with respect to $J$
and $N$ is shown at $\Rm=10^5$ for $\v_2$ in Table~1. The higher order modes require
even
greater truncation levels, as they exhibit steeper spatial gradients (being
more strongly localised). As $\Rm$ is
increased further ($\Rm\!>\!5\!\times\!10^5$) difficulties are encountered
 because the truncation needed (and consequently the size of the
matrices generated) become prohibitive.

\subsection{Growth rates}

We present below the growth rates of the $n=0,1,2$ magnetic modes for $k=1$ and
$m=1$ in the two
spherical
helical dynamos we considered.

\begin{table}
\begin{center}
\begin{footnotesize}
\begin{tabular}{r|rr|rr|rr}
\hline
$\Rm$ & $\Re\lambda_0$ &  $\Im\lambda_0$ &  $\Re\lambda_1$ & $\Im\lambda_1$ &
$\Re\lambda_2$ & $\Im\lambda_2$ \\
\hline
    500 &  17.8 &    18.2 & $-60.3$ &    17.8 &       -- &      -- \\
  1,000 &  38.0 &    74.7 & $-48.0$ &    66.1 & $-642.4$ &   113.6 \\
  2,000 &  68.9 &   206.3 & $-25.0$ &   187.4 & $-384.8$ &   196.7 \\
  3,000 &  93.6 &   347.2 &  $-5.2$ &   320.8 & $-375.0$ &   322.2 \\
  5,000 & 133.2 &   641.3 &  $27.4$ &   602.2 & $-348.1$ &   592.7 \\
 10,000 & 203.8 &  1408.9 &  $87.8$ &  1344.3 & $-282.1$ &  1303.8 \\
 20,000 & 293.9 &  3006.6 & $178.5$ &  2878.7 & $-211.5$ &  2747.8 \\
 30,000 & 363.0 &  4645.6 & $224.4$ &  4423.7 & $-237.0$ &  4339.7 \\
 50,000 & 477.1 &  7969.4 & $250.7$ &  7697.2 &  $-86.1$ &  7546.4 \\
100,000 & 687.5 & 16392.3 & $348.9$ & 16000.0 &  $-19.2$ & 15566.2 \\
200,000 & 982.4 & 33432.1 & $486.2$ & 32867.1 &  $-57.2$ & 32299.4 \\
\hline
\end{tabular}
\end{footnotesize}
\end{center}
\caption{The growth rates of the leading modes as computed by the numerical
eigenproblem
for $\v_2$, $m=1$ at different $\Rm$. The subscript on $\lambda$ indicates the mode
number $n$. Numerical truncation levels are $N=40$ and $J=800$}
\end{table}

 Table~2 shows
the growth
rates of the $n=0,1,2$ modes, as computed by the numerical eigenproblem for
$\v_2$ at the truncation levels $N=40$ and $J=800$. For sufficiently large
$\Rm$, 
there exist two
growing dynamo modes corresponding to $n=0$ and $n=1$. Note
the scaling
$\Im\lambda/\Re\lambda\sim \Rm^{1/2}$ predicted by the asymptotic theory for
large $\Rm$
(see Equation \eqref{2.18}).
 Moreover, the leading modes
possess
growth rates whose imaginary parts asymptote to a common value (see Equation
\eqref{2.58}).
This characteristic clustering of the eigenvalues in spherical helical dynamos
explains the
difficulty that algebraic eigensolvers encounter when separating the eigenvalues at large
$\Rm$. In
this regime a partial eigensolver, such as the implicitly restarted Arnoldi
method, is
invaluable (see Latter and Ivers 2004).

\begin{figure}
\begin{center}
\scalebox{0.5}{\includegraphics{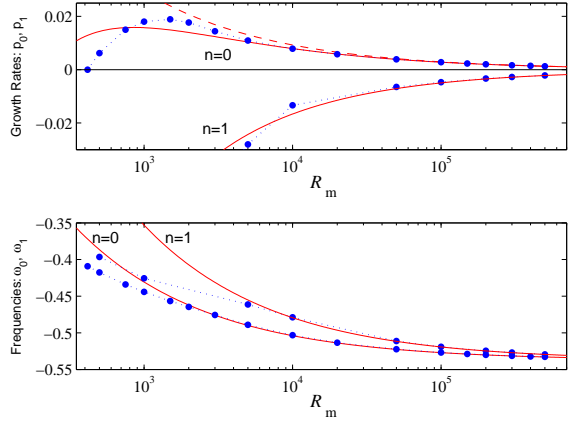}}
\end{center}
\caption{The growth rates and angular frequencies of the two leading modes of the
  $\v_1$ flow. These are characterised by $m=1$,
$k=1$, and
$n=0$ or $n=1$. The solid line represents the asymptotic expressions with the extra
terms of
Gilbert \& Ponty (2000), the dashed line represents the asymptotic expression correct to order
$\varepsilon^2$, the points on the dotted line represent the full numerical
eigensolution.}
\end{figure}

\begin{figure}
\begin{center}
\scalebox{0.5}{\includegraphics{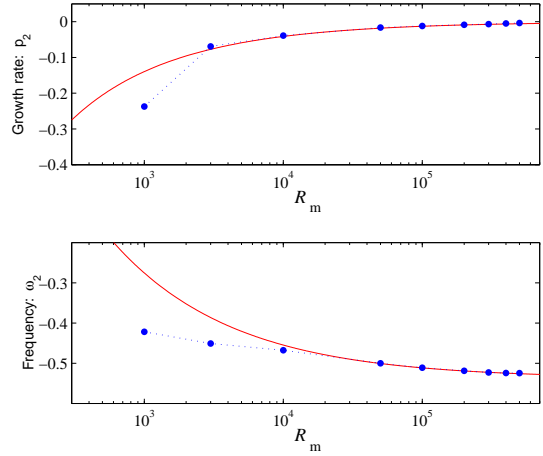}}
\end{center}
\caption{The growth rate $p_2$ and angular frequency $\omega_2$ for the $m=1$,
$k=1$,
$n=2$ mode of $\v_1$.}
\end{figure}

\begin{figure}
\begin{center}
\scalebox{0.5}{\includegraphics{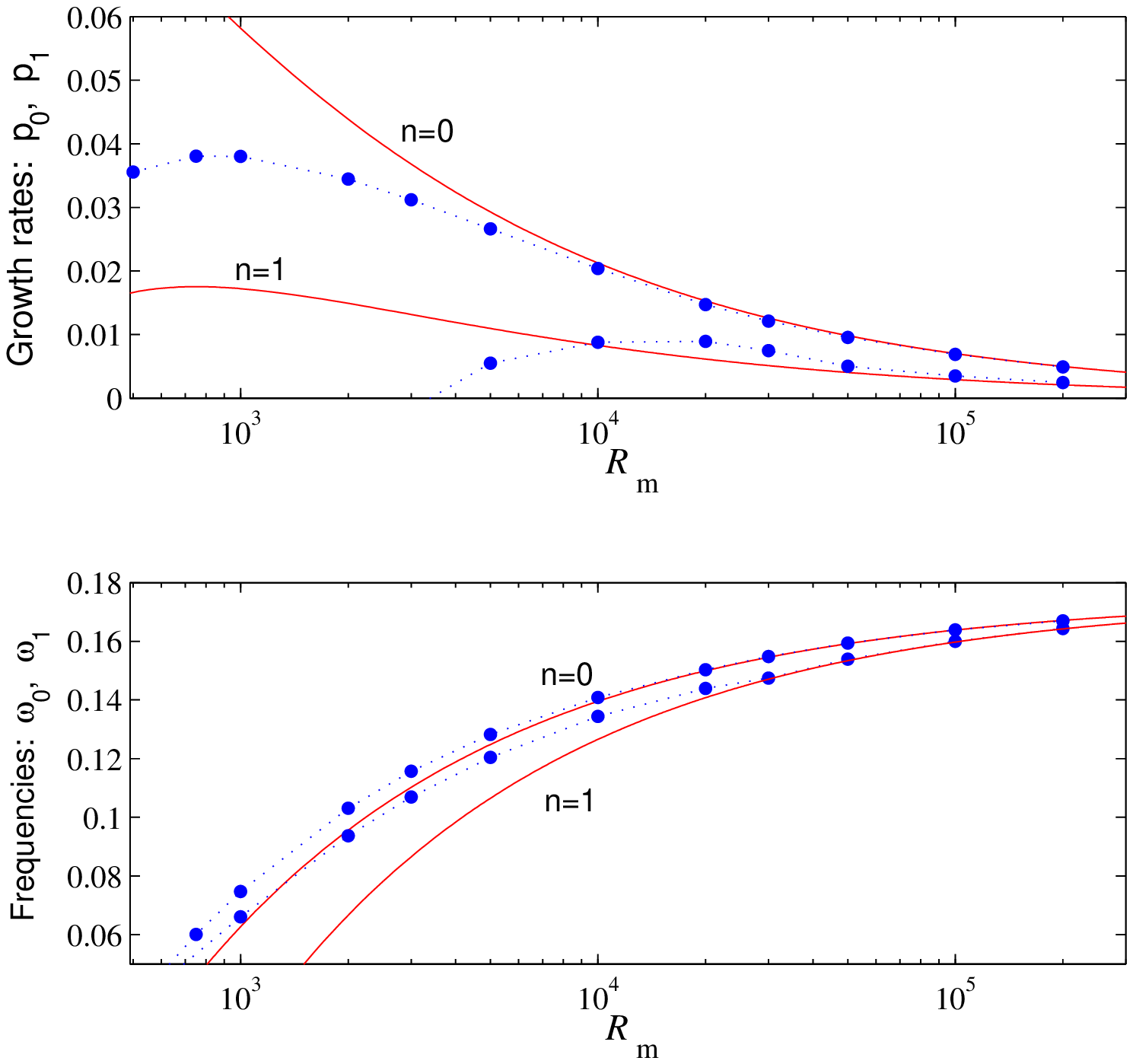}}
\end{center}
\caption{The growth rates and angular frequencies of the leading two modes of
  the $\v_2$ dynamo. Here $m=1$, $k=1$ and
$n=0$ or $n=1$.}
\end{figure}

In Figs 3 and 4 we directly compare the predictions of
the asymptotic theory and the numerical computations of the full eigenproblem
for $\v_1$.
Here is presented the numerical growth rate $\text{Re}\,\lambda$, and associated 
angular frequency
$\text{Im}\,\lambda$, of the leading modes as a function of
$\Rm$. Alongside these data points, we plot the asymptotic growth rates $p_i$ 
and frequencies $\omega_i$,
calculated with (a) the higher order terms
 of GP  included (the solid
lines), (b) the asymptotic theory correct to order
$\varepsilon^2$
(the dashed line).
The truncation
levels for
the asymptotic values are $K=400$. Note that both growth rates are scaled on
the turnover time. This makes clear that the numerical growth rate goes to
zero as $\Rm\to \infty$. Helical dynamos are `slow', as expected 
(Childress and Gilbert 1995, GP). Of the
three modes shown, only the $n=0$ mode is a dynamo, which is active above a critical
magnetic
Reynolds
number of $\Rm\approx 416$ and achieves its maximum positive growth rate at
$\Rm\approx 1500$.
In Fig.~5 we plot the growth rates and frequencies of the leading two modes of
the $\v_2$ dynamo, for $m=k=1$. Both of these modes may grow for sufficiently
large $\Rm$.

Plainly, there is excellent agreement between the $n=0$ asymptotic growth rates
and the numerical growth rates when $\Rm\!\gtrsim\!10^4$ for both $\v_1$ and
 $\v_2$. The angular frequencies agree at smaller $\Rm$.
Fig.3 also shows that the additional terms of GP at
higher order $\varepsilon^4$ improve the accuracy of the asymptotics markedly,
in comparison with the theory up to $\varepsilon^2$ (the dashed line).
 This agreement strongly supports the asymptotic theory. It also
indicates that
the identification of the numerical modes with the asymptotic modes is
correct. Finally, this shows that these flows only admit Ponomarenko-type
dynamos in the large $\Rm$ regime.

\begin{figure}
\begin{center}
\scalebox{0.85}{\includegraphics{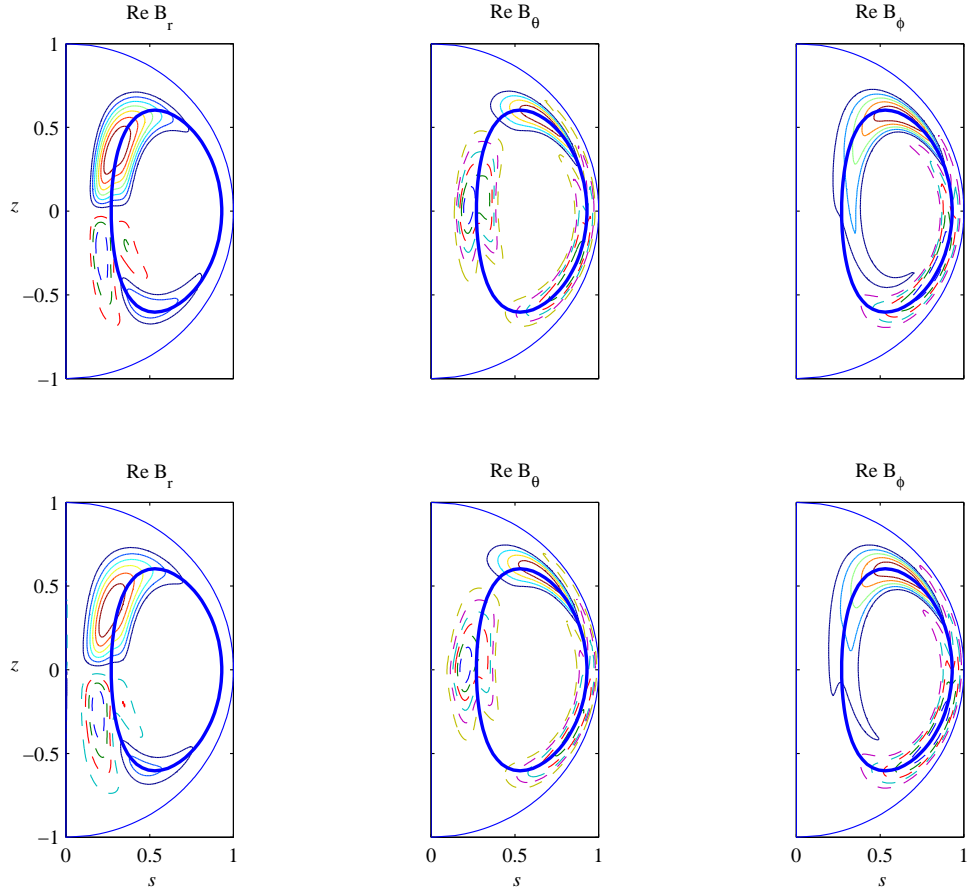}}
\caption{Plots of the magnetic eigenfunctions: the real parts of ${\hat B}_r$, ${\hat
B}_\theta$,
${\hat B}_\phi$, for the $m=1$, $k=1$, $n=0$ mode upon the
$\v_1$ flow.
Here
$\sigma=\sigma_1(1,1,\Psi_0)$ and
$\Rm=10^5$. The three upper panels present the numerical eigensolution,
while the lower three panels present the asymptotic eigensolution. Both
numerical and asymptotic
 eigenfunctions have been normalised so that $|\hat\B|=1$, and each component has
 been plotted with 10 contours of equal increments.}
\end{center}
\end{figure}

\begin{figure}
\begin{center}
\scalebox{0.85}{\includegraphics{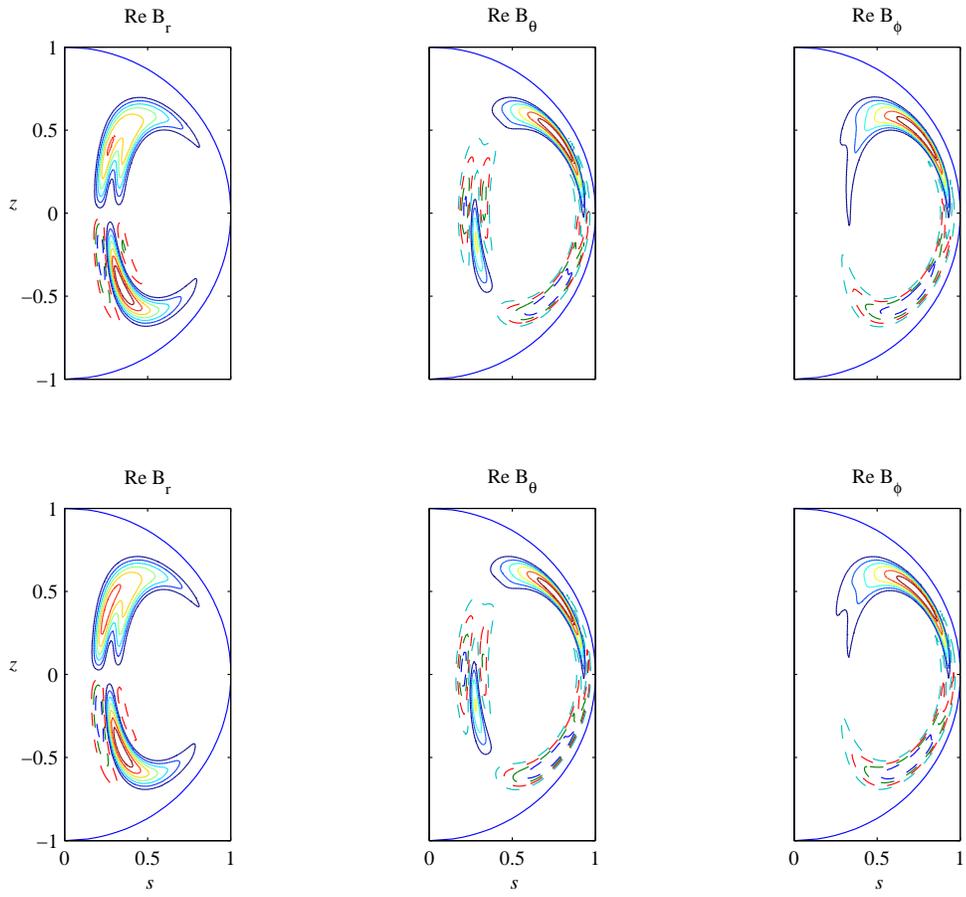}}
\caption{Plots of the real parts of the magnetic eigenfunctions: as in Figure 6 but with
  $\Rm=5\times 10^5$.}
\end{center}
\end{figure}

\begin{figure}
\begin{center}
\scalebox{0.85}{\includegraphics{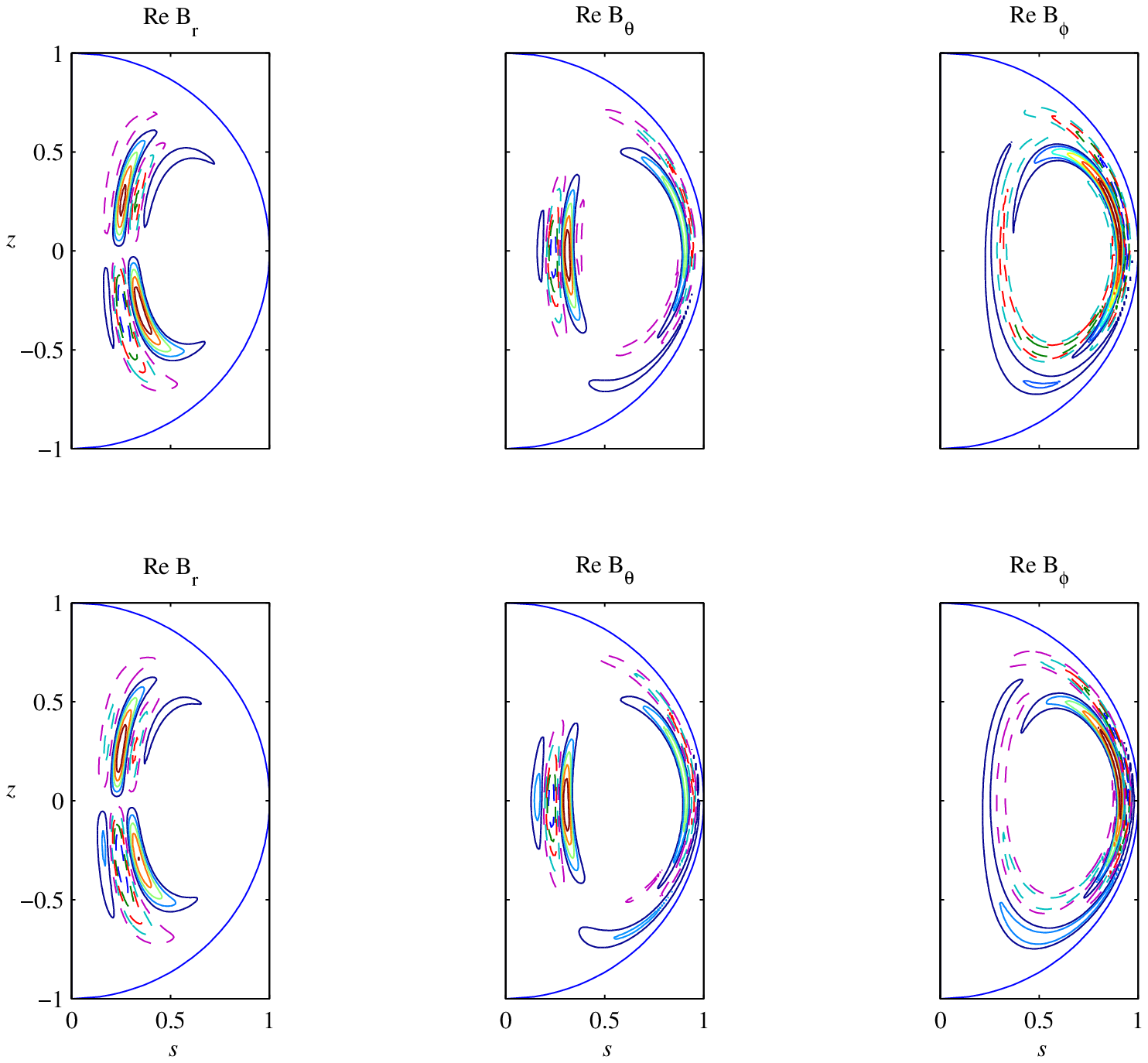}}
\caption{Plots of the magnetic eigenfunctions: the real part of the
  $n=1$ eigenmode, with $k=m=1$ and $\Rm=5\times 10^5$ upon flow $\v_1$..}
\end{center}
\end{figure}

\subsection{Magnetic field structure}

Figures~6--8 show the real parts of the magnetic eigenfunctions for different
values of $\Rm$ and $n$ upon the $\v_1$ flow. These choices reveal the salient physical
and asymptotic features of these modes. The morphologies of the $\v_2$ dynamos
are much the same and are omitted in the interests of space.

In Fig.~6 we plot the
 magnetic field components, ${\hat B}_r$, ${\hat
B}_\theta$,
${\hat B}_\phi$, of the $m=1$, $k=1$, $n=0$ mode.
Here
$\sigma=\sigma_1(1,1,\Psi_0)$ and we set the magnetic Reynolds number to
$\Rm=10^5$. The top three panels present the numerical
eigensolutions, while the bottom three panels show the asymptotic
approximations. Superimposed upon the first set
 of figures is the resonant stream curve,
 which helps
highlight
the localisation of the magnetic field. 

It is apparent from the figures that
the
asymptotic and numerical magnetic fields agree in their dominant features: the
position,
orientation and shape of the local maxima and minima.
 The localisation of the field to the resonant streamline is
readily
observed, with marked flux expulsion inside and outside the resonant
streamline, as expected.
 In addition, the $k=1$ nature of the field is clear from its
variation
around the streamline, especially in $\hat{B}_\phi$.
 Note, however, that there is a slight offset
outwards
away from the resonant streamline in the upper numerical eigenfunctions. 
This offset in combination with the steep spatial gradients means the relative
error between the asymptotic and numerical eigenfunctions is larger than
expected. The relative error, defined by $\|
\B^\text{num}-\B^\text{asym}\|_2/\|\B^\text{num}\|_2$, is 0.334 for this
$\Rm=10^5$ case.

In Fig.~7 we show the same mode but for larger magnetic Reynolds number:
$\Rm=5\times10^5$. The resonant stream curve has been omitted.
 The magnetic field is more
localised and intense, with the offset of the maxima
and minima
from the resonant streamline substantially reduced.
 Now the magnitudes of the numerical and
asymptotic eigenfunction, in addition to the orientation of
the magnetic features, are in good agreement. The relative error
 is reduced to 0.285.

The more involved $n=1$ eigenmode is plotted in Figure 8.
 The other parameters
are kept the same and $\Rm=5\times10^5$. 
 This mode decays slowly with time
as shown
in Figure~3. Its structure is more complicated due to the variation
with $\psi$
under the gaussian envelope of $D_1(\Upsilon/\kappa)$. In particular, the field
components vanish on the resonant streamline. The agreement between the
asymptotic and numerical eigenfunctions is good but not as striking as in the
$n=0$ case. In particular, the relative magnitudes of the maxima and minima
are not in agreement, though the general shape of the structures are.
 This is possibly due to the fact that the mode has not fully
converged to its asymptotic form. On the other hand, its greater spatial
variation may be taxing the resolution of our numerical scheme, leading to errors.

\section{Conclusions}

In this paper we have compared 
the asymptotic theory of Gilbert \& Ponty (2000) for axisymmetric roll dynamos
in a
sphere to the numerically computed results of the exact
dynamo problem
for two simple flows, with azimuthal components of the special form
$v_\phi=r\sin\theta\,W(\psi)$ and of a more general form. In the regime $\Rm\gtrsim
10^4$ excellent agreement is obtained
between
the asymptotic theory to $\mathcal{O}(\Rm^{-1/2})$ and the numerical results
for the
growth rate and angular frequency. For
the
magnetic field the agreement between the asymptotic theory at leading order and
the
numerical results is good if $\Rm=10^5$ and excellent if $\Rm=5\!\times\!
10^5$. The
asymptotic formulas for the growth rate and the angular frequency have been
extended in Appendix A in the Supplemental Material.

Only the simplest class of axisymmetric roll dynamos have been considered:
those which
consist of a single-roll flow with a single resonant streamline. The magnetic
field in
these dynamos is localised to the resonant stream surface and can interact only
with
itself. Further work is required on more complicated spherical roll flows:
such as those
those with a
single-roll but more than one resonant streamline, and those with several
rolls.
These
flows offer the possibility of interaction between magnetic fields localised to
separate
regions of the flow. This may produce interacting modes of non-Ponomarenko
type, e.g.
Gailitis type modes, alongside the Ponomarenko type modes 
(Gailitis 1970, 1993, 1995, Moss 2006). A related question,
which arises from the localised nature of the Ponomarenko modes, is
whether they depend on the magnetic boundary conditions at the surface of the
conducting
fluid. 

Of key interest to both laboratory and astrophysical 
applications are (a) the nonlinear saturation of
 such modes and (b) their relationship to a background of small-scale velocity
 fluctuations.
How will the dynamo modes back react on the
helical flow which generated them? Will these modes build up magnetic torques
which stifle the meridional motion, or will a more complicated dynamical 
interplay arise? If fluctuations are indeed present, for what
size amplitudes, and for what correlation times and lengths,
 will they succesfully impede the satisfaction of the resonance
 condition which is so crucial for the formation of Ponomarenko modes? Recent
 work in cylindrical geometry for time-dependent fluctuations shows that
 Ponamarenko-type dynamo action can, in fact, survive in certain cases
 (Peyrot et al.\ 2007, 2008). But this need not be the case
 at all for flows exhibiting small-scale variations in space, and 
it is the latter situation which is probably most relevant in applications.

\section*{Acknowledgments}

The authors wish to thank the two anonymous referees for thorough and
helpful reviews. 
Their comments greatly improved the manuscript. HNL acknowledges funding from the
University of Sydney via a University Postgraduate Award.

\newpage

\newpage

\begin{center}
\begin{Huge}
Spherical single-roll dynamos at large magnetic Reynolds number: Supplemental material
\end{Huge}
\end{center}

\appendix
\section{Derivation of asymptotic expressions at large $\Rm$}

In this appendix we briefly derive the leading order results of the GP asymptotic
theory presented in Section 3 of the main manuscript `Spherical single-roll
dynamocs at large magnetic Reynolds number'. Once these are given we sketch out a technique
whereby higher order terms may be calculated and give the solution to order
$\varepsilon^3$. Note that reference labels for equations in the main
manuscript are not preceded by either an `A' or `B'.

\subsection{Preliminaries}

\subsubsection{The magnetic diffusion term in toroidal coordinates}

We take as our starting point the magnetic induction equation in the toroidal
coordinate system $(\psi,\,\vartheta,\,\zeta)$, Eqs
\eqref{2.15}-\eqref{2.17} in the main manuscript.
 In order to progress, the diffusion terms on the
right sides need to be decomposed into their component parts. 

Using the summation convention the magnetic field can be written as $\B=B^i\f_i$,
the gradient
operator as $\nabla=(\nabla\psi^i)\d_i$, where $\d_i=\d_{\psi^i}$, and the
diffusion term in
the magnetic induction equation as $\nabla^2\B = (\nabla^2B^j)\f_j + 2\nabla
B^j\cdot\nabla\f_j
+ B^j\nabla^2\f_j$. Thus, since $\nabla B^j=(\d_kB^j)\nabla\psi^k$, the covariant
components of this term are
\[
\nabla\psi^i\cdot\nabla^2\B = (\nabla^2B^i) + 2(\d_kB^j)
\nabla\psi^k\cdot\nabla\f_j\cdot\nabla\psi^i +
B^j\nabla\psi^i\cdot\nabla^2\f_j\,.
\]
Four of the 27 terms $\nabla\psi^k\cdot\nabla\f_j\cdot\nabla\psi^i$ and three of the
9 terms
$\nabla\psi^i\cdot\nabla^2\f_j$ vanish identically, since
$\grad\f_\zeta=\1_s\1_\phi-\1_\phi\1_s$ means that $\a\cdot(\grad\f_\zeta)\cdot\b=0$
for any
meridional vectors $\a$, $\b$. We then have
\[
\nabla\psi\cdot(\nabla\psi\cdot\nabla\f_\zeta) =
\nabla\psi\cdot(\nabla\vartheta\cdot\nabla\f_\zeta) =
\nabla\vartheta\cdot(\nabla\psi\cdot\nabla\f_\zeta) =
\nabla\vartheta\cdot(\nabla\vartheta\cdot\nabla\f_\zeta) = 0\,.
\]
Also $\nabla^2\f_\zeta=\0$ implies $\nabla\psi\cdot\nabla^2\f_\zeta =
\nabla\vartheta\cdot\nabla^2\f_\zeta = \nabla\zeta\cdot\nabla^2\f_\zeta = 0$.

Keeping only terms which appear later in the asymptotic analysis and suppressing the
others
with dots, the relevant diffusion terms are
\begin{align}
  \nabla\psi\cdot\nabla^2\B &= (\nabla^2 + 2\mu_i\d_\psi + 2\mu_j\d_\vartheta +
2\mu_k\d_\zeta+ \mu_l)B_\psi \notag \\
& \hskip3cm  + (2\mu_a\d_\psi + 2\mu_b\d_\vartheta + 2\mu_c\d_\zeta + \mu_d)B_\vartheta
  + 2\mu_g\d_\zeta B_\zeta
  \label{2.10}\\
  \nabla\vartheta\cdot \nabla^2\B &=
  ( \nabla^2 + 2\lambda_a\d_\psi + 2\lambda_b\d_\vartheta + 2\lambda_c\d_\zeta
+ \lambda_d)B_\vartheta
  + 2\lambda_g\d_\zeta B_\zeta + \dots
  \label{2.11}\\
  \nabla\zeta\cdot \nabla^2\B &=  2\rho_a\d_\psi B_\vartheta
  + ( \nabla^2 + 2\rho_b\d_\psi ) B_\zeta + \dots
  \label{2.12}\,,
\end{align}
where the coefficients are defined by
\begin{align*}
  \mu_a &= \nabla\psi\cdot(\nabla\psi\cdot\nabla\f_\vartheta) &
  \mu_b &= \nabla\psi\cdot(\nabla\vartheta\cdot\nabla\f_\vartheta) &
  \mu_c &= \nabla\psi\cdot(\nabla\zeta\cdot\nabla\f_\vartheta) &
  \mu_d &= \nabla\psi\cdot\nabla^2\f_\vartheta \\
  \mu_g &= \nabla\psi\cdot(\nabla\zeta\cdot\nabla\f_\zeta) &
  \mu_i &= \nabla\psi\cdot(\nabla\psi\cdot\nabla\f_\psi) &
  \mu_j &= \nabla\psi\cdot(\nabla\vartheta\cdot\nabla\f_\psi) &
  \mu_k &= \nabla\psi\cdot(\nabla\zeta\cdot\nabla\f_\psi) \\
  \mu_l &= \nabla\psi\cdot\nabla^2\f_\psi &
  \lambda_a &= \nabla\vartheta\cdot(\nabla\psi\cdot\nabla\f_\vartheta) &
  \lambda_b &= \nabla\vartheta\cdot(\nabla\vartheta\cdot\nabla\f_\vartheta) &
  \lambda_c &= \nabla\vartheta\cdot(\nabla\zeta\cdot\nabla\f_\vartheta) \\
  \lambda_d &= \nabla\vartheta\cdot\nabla^2\f_\vartheta &
  \lambda_g &= \nabla\vartheta\cdot(\nabla\zeta\cdot\nabla\f_\zeta) &
  \rho_a &= \nabla\zeta\cdot(\nabla\psi\cdot\nabla\f_\vartheta) &
  \rho_b &= \nabla\zeta\cdot(\nabla\psi\cdot\nabla\f_\zeta)\,.
\end{align*}
Apart from $\mu_d$, $\mu_l$, $\lambda_d$ these are Christoffel symbols.

The scalar Laplacian is $\nabla^2 = \nabla\cdot(\nabla\psi^i)\d_i =
(\nabla\psi^i\cdot\nabla\psi^j)\d_i\d_j + (\nabla^2\psi^i)\d_i$, which, in full, can be
expressed as
\begin{multline*}
\nabla^2 = (\nabla\psi)^2\d^2_\psi +
2(\nabla\psi\cdot\nabla\vartheta)\d_\psi\d_\vartheta
+ (\nabla\vartheta)^2\d^2_\vartheta +
2(\nabla\vartheta\cdot\nabla\zeta)\d_\vartheta\d_\zeta  \\
+ (\nabla\zeta)^2\d^2_\zeta + 2(\nabla\zeta\cdot\nabla\psi)\d_\zeta\d_\psi +
(\nabla^2\psi)\d_\psi + (\nabla^2\vartheta)\d_\vartheta +
(\nabla^2\zeta)\d_\zeta\,.
\end{multline*}
Furthermore, a number of the geometric coefficients average to zero,
\begin{equation} \label{2.13}
  \overline{\mu}_a = \overline{\mu}_d = \overline{\mu}_g = \overline{\mu}_l =
\overline{\lambda}_b = \overline{\rho}_b = 0\,,
\end{equation}
which can be established using standard vector
identities, the
divergence theorem and Stokes' theorem (see GP).

\subsubsection{Explicit asymptotic expansions}

We now employ the asymptotic scalings presented in Section 3.2 and the order
one variable $\varUpsilon$, and then choose a
specific streamsurface $\psi=\psi_0$ around which we expand.

The $\psi$-derivatives and
gradient
operators in Eqs \eqref{2.15}-\eqref{2.17} become
\begin{equation} \label{2.20}
  \d_\psi = \varepsilon^{-1}\d_\varUpsilon\,,\qquad
  \nabla = \varepsilon^{-1} \nabla\psi\,\d_\varUpsilon +
\nabla\vartheta\,\d_\vartheta + \nabla\zeta\,\d_\zeta\,,
\end{equation}
and the magnetic field components take the functional forms,
\begin{equation} \label{2.21}
B_\psi = \varepsilon^2 b_\psi(\varUpsilon,\vartheta) e^{im\zeta + i\omega t +
pt}\,,\quad
B_\vartheta = b_\vartheta(\varUpsilon,\vartheta) e^{im\zeta + i\omega t + pt}\,,\quad
B_\zeta = b_\zeta(\varUpsilon,\vartheta) e^{im\zeta + i\omega t + pt}\,.
\end{equation}

We now expand $\omega$ and $p$ in powers of $\varepsilon$ with the ordering \eq{2.18},
\begin{equation}\label{2.22}
   \omega = \omega_0 + \varepsilon\omega_1 + \varepsilon^2\omega_2
   + \varepsilon^3\omega_3 + \varepsilon^4\omega_5 + \dots\,,\quad
   p = \varepsilon^2p_2 + \varepsilon^3p_3  + \varepsilon^4p_4 + \dots\,,
\end{equation}
and expand $\overline{W}(\psi)$ and $\Omega(\psi)$ in Taylor series about the
streamline
$\psi=\psi_o$,
\begin{align}
  \Omega(\psi_o + \varepsilon\varUpsilon) &=
  \Omega_o + \Omega'_o\varepsilon\varUpsilon +
\tfrac{1}{2}\Omega''_o\varepsilon^2\varUpsilon^2 + \dots
  \label{2.23}\\
  \overline{W}(\psi_o + \varepsilon\varUpsilon) &=
  \overline{W}_o + \overline{W}'_o\varepsilon\varUpsilon +
\tfrac{1}{2}\overline{W}''_o\varepsilon^2\varUpsilon^2 + \dots\,,
  \label{2.24}
\end{align}
in which $\Omega_o = \Omega(\psi_o)$, $\Omega'_o = \Omega'(\psi_o)$, etc. Assuming the
functional dependencies of \eq{2.21} and substituting the expansions
\eq{2.22}--\eq{2.24} into
the advection operator \eq{2.5} gives
\begin{equation} \label{2.25}
  D_t = d_0 + \varepsilon d_1 + \varepsilon^2 d_2 + \varepsilon^3 d_3 +
\varepsilon^4 d_4 + \dots\,,
\end{equation}
where
\begin{equation}\label{2.26}
  d_n = p_n + i\omega_n + \frac{\varUpsilon^n}{n!}\Big(\Omega^{(n)}_o\d_\vartheta
  + im\overline{W}^{(n)}_o\Big)\,,\quad p_0=p_1=0\,.
\end{equation}
We also expand $(\nabla\psi)^2$ in the diffusion term,
\begin{equation} \label{2.31}
  (\nabla\psi)^2 = \gamma_0 + \varepsilon\varUpsilon \gamma_1
  + \varepsilon^2\varUpsilon^2 \gamma_2 + \varepsilon^3\varUpsilon^3 \gamma_3
  + \varepsilon^4\varUpsilon^4 \gamma_4 + \dots\,
\end{equation}
as well as the individual `diffusion coefficients', for example:
$$\mu_b=\mu_{b,0} + \varepsilon\varUpsilon\mu_{b,1}+\dots.$$
Finally we expand the magnetic field components,
\begin{equation}\label{2.32}
  b_\psi = b_{\psi 0} + \varepsilon b_{\psi 1} + \dots\,,\quad
  b_\vartheta = b_{\vartheta 0} + \varepsilon b_{\vartheta 1} + \dots\,,\quad
  b_{\zeta} = b_{\zeta 0} + \varepsilon b_{\zeta 1} + \dots\,.
\end{equation}
We are now ready to derive the asymptotic equations at the various orders.

\subsection{The $\varepsilon^0$ equations}

In this and the following two subsections we describe the asymptotics to order
 $\varepsilon^2$.
 We substitute expansions \eq{2.22}--\eq{2.25}, \eq{2.31}
and \eq{2.32}
into the component equations \eq{2.15}--\eq{2.17}, divide \eq{2.15} by
$\varepsilon^2$, and
collect terms of like order.

The $\varepsilon^0$-equations are
\begin{equation} \label{2.33}
  d_0 b_{\psi 0} = 0\,,\qquad d_0 b_{\vartheta 0} = 0\,,\qquad d_0b_{\zeta 0} = 0\,,
\end{equation}
which have the solution,
\begin{equation} \label{2.34}
 b_{\psi 0} = F_{\psi 0}(\varUpsilon)e^{ik\vartheta}\,,\quad
 b_{\vartheta 0} = F_{\vartheta 0}(\varUpsilon)e^{ik\vartheta}\,,\quad
 b_{\zeta 0} = F_{\zeta 0}(\varUpsilon)e^{ik\vartheta}\,,
\end{equation}
where the functions $F_{\psi 0}$, $F_{\vartheta 0}$, $F_{\zeta 0}$ are determined at
order
$\varepsilon^2$ and must vanish as $|\varUpsilon|\to\infty$. The constant $k$ is an
integer
since $\B$ is single-valued. Solvability of \eq{2.33} fixes the angular frequency
$\omega$ to
leading order for given $m$ and $k$,
\begin{equation} \label{2.35}
  \omega_0 = -\Pi_o\,,
\end{equation}
where we have introduced the advection frequency function $\Pi(\psi)=k\Omega +
m\overline{W}$. The
operator $d_0$ becomes $\Omega_o(\d_\vartheta-ik)$, and hence annihilates any term
with the
$\vartheta$-dependence $e^{ik\vartheta}$. 

\subsection{The $\varepsilon^1$ equations}

The $\varepsilon^1$-equations are
\begin{equation}\label{2.36}
  d_0 b_{\psi 1} + d_1 b_{\psi 0} = 2\mu_{a,o}\d_\varUpsilon b_{\vartheta 0}\,,\quad
  d_0 b_{\vartheta 1} +  d_1 b_{\vartheta 0} = 0\,,\quad
  d_0 b_{\zeta 1} + d_1 b_{\zeta 0} = 0\,.
\end{equation}
Their solvability requires
\begin{equation} \label{2.37}
  \omega_1 = 0\,,\qquad \Pi'_o = k\Omega'_o + m\overline{W}'_o = 0\,.
\end{equation}
The last condition fixes the resonant streamline $\psi= \psi_o$, upon which the
magnetic field
is localised for given $m$ and $k$. At this streamline the function $\Pi(\psi)$
possesses a
critical point, and a maximum if $\Pi''_o<0$, which is the case for the simple roll
flows we
examine. The larger gradients in $B_\vartheta$ and $B_\zeta$ on this surface encourage
diffusion of these fields and hence replenishment of $B_\psi$. The operator $d_1$
becomes
$d_1=\Upsilon\Omega'_o(\d_\vartheta-ik)$ and hence also annihilates any term with
$\vartheta$-dependence $e^{ik\vartheta}$.

The last two equations in \eq{2.36} can be solved similarly to \eq{2.33}. The first
equation
reduces to,
\[
  d_0 b_{\psi 1} = 2 \mu_{a,o} \d_\varUpsilon b_{\vartheta 0} = 2\mu_{a,o}
  F'_{\vartheta 0}(\varUpsilon) e^{ik\vartheta}\,,
\]
which is solvable, since $\overline{\mu}_a=0$ and the right side then possesses no
term with
the $\vartheta$-dependence $e^{ik\vartheta}$. Thus the magnetic field at order
$\varepsilon^1$
is
\begin{align}\label{2.38}
  b_{\psi 1} = F_{\psi 1}(\varUpsilon)e^{ik\vartheta} + G_{\psi
1}(\varUpsilon,\vartheta)e^{ik\vartheta}\,,\quad
  b_{\vartheta 1} = F_{\vartheta 1}(\varUpsilon)e^{ik\vartheta}\,,\quad
  b_{\zeta 1} = F_{\zeta 1}(\varUpsilon) e^{ik\vartheta}\,,
\end{align}
where the functions $F_{\psi 1}$, $F_{\vartheta 1}$, $F_{\zeta 1}$ are determined at
order
$\varepsilon^3$ and the particular integral for equation \eq{2.36}(a) is
\begin{equation}\label{2.39}
  G_{\psi 1} = \frac{2 F'_{\vartheta 0}}{\Omega_o}\widehat{\mu}_{a,o}\,,\qquad
  \overline{G_{\psi 1}} = 0\,.
\end{equation}
Here we have introduced the hat operator $\ \widehat{}\ $ defined by
\[ \d_\vartheta \widehat{f} \equiv f - \overline{f}\,,\qquad \overline{\!\widehat
f\:}=0\,, \]
which implies
\[
\widehat{f} = \int_0^\vartheta (f - \overline{f})\,d\vartheta
- \overline{\int_0^\vartheta (f - \overline{f})\,d\vartheta}\,.
\]
In addition, the properties
\begin{equation}\label{2.39.0}
    \overline{\widehat{f}g} = - \overline{f\widehat{g}\:}\,,\qquad
\overline{f\widehat{f}\:}=0
\end{equation}
are easily established.

\subsection{The $\varepsilon^2$ equations}

The $\varepsilon^2$-equations are
\begin{align}
  d_0 b_{\psi 2} + d_1 b_{\psi 1} + (d_2 - \gamma_0\d^2_\varUpsilon) b_{\psi 0} &=
  2\mu_{a,o}\d_\varUpsilon b_{\vartheta 1} + 2\mu'_{a,o}\varUpsilon\d_\varUpsilon
b_{\vartheta 0}
  \nonumber\\
  & + (2\mu_{b,o}\d_\vartheta + 2im\mu_{c,o} + \mu_{d,o}) b_{\vartheta 0}
  + 2im\mu_{g,o} b_{\zeta 0}
  \label{2.40}\\
  d_0 b_{\vartheta 2} + d_1 b_{\vartheta 1} + (d_2 - \gamma_0\d^2_\varUpsilon)
b_{\vartheta 0} &=
  \Omega'_o b_{\psi 0}
  \label{2.41}\\
  d_0 b_{\zeta 2} + d_1 b_{\zeta 1} + (d_2 - \gamma_0\d^2_\varUpsilon) b_{\zeta 0} &=
  \overline{W}'_o b_{\psi 0}\,.
  \label{2.42}
\end{align}
GP included subdominant terms from the Laplacian at this order
arguing
that these are comparable when employing the scalings of  Gilbert (1988). However,
they neglect
to include the coordinate Laplacians, $\nabla^2\psi$, $\nabla^2\vartheta$,
$\nabla^2\zeta$
which should be of the same order. In the present analysis all these terms appear at the
correct
(higher) orders.

Equations \eq{2.40}--\eq{2.42} are solvable for the field components $b_{\psi 2}$,
$b_{\vartheta 2}$ and $b_{\zeta 2}$, if the $\vartheta$-dependence of the other
terms is not
$e^{ik\vartheta}$. This is true for the terms, $d_1 b_{\psi 1}$, $d_1 b_{\vartheta
1}$ and $d_1
b_{\zeta 1}$, since the operator $d_1$ annihilates $e^{ik\vartheta}$. It is also
satisfied by
the terms, $2\mu_{a,o}\d_\varUpsilon b_{\vartheta 1}$, $2\mu_{d,o} b_{\vartheta 0}$ and
$2im\mu_{g,o} b_{\zeta 0}$, since the coefficients $\mu_{a,o}$, $\mu_{d,o}$,
$\mu_{g,o}$
average to zero by \eq{2.13}. The sum of the remaining terms must average to zero after
multiplication by $e^{-ik\vartheta}$. Consequently we may write the solvability
condition for
\eq{2.40}--\eq{2.42} as
\begin{align}\label{2.43}
\mathbf{L}\left( \begin{matrix}
F_{\psi 0} \\
F_{\vartheta 0} \\
F_{\zeta 0}
\end{matrix} \right)= \0\,, \qquad
\mathbf{L} = \left( \begin{matrix}
\Xi & 2i\overline{\alpha}_o & 0 \\
\Omega'_o & \Xi & 0 \\
\overline{W}'_o & 0 & \Xi
\end{matrix} \right)\,,
\end{align}
where
\begin{equation}\label{2.44}
  \Xi =  \overline{\gamma}_0 \d_\varUpsilon^2 -
\tfrac{1}{2}i\Pi''_o\varUpsilon^2 - p_2 - i\omega_2\,,\qquad
  \alpha = k\mu_b + m\mu_c\,,\qquad
  \Pi''_o  = k\Omega''_o + m\overline{W}''_o\,,
\end{equation}
with $\overline{\gamma}_0 = <|\nabla\psi|_o^2>$. These equations determine the
functions
$F_{\psi 0}$, $F_{\vartheta 0}$, $F_{\zeta 0}$ and hence the magnetic field to
leading order.
The solutions are of the form
\begin{equation} \label{2.45}
  \left( \begin{matrix}
  F_{\psi 0} \\
  F_{\vartheta 0} \\
  F_{\zeta 0}
  \end{matrix}\right) =
  y_n(\varUpsilon)\mathbf{a}\,,\qquad
  y_n(\varUpsilon) = \D_n\left( \varUpsilon/\kappa\right), \qquad
  \kappa = (\overline{\gamma}_0/2i\Pi''_o)^{1/4}.
\end{equation}
Here $\mathbf{a}$ is a constant vector to be determined and $\D_n(z)$ is the
parabolic cylinder
function of degree $n$,
\[ \D_n(z) = 2^{-n/2}e^{-z^2/4}\He_n(z/\sqrt{2})\,,\qquad n\geq 0\,, \]
where $\He_n(z)$ is the Hermite polynomial of degree $n$ (Abramowitz \& Stegun
1972). 
In order for this solution to satisfy the boundary conditions, i.e. $F_{\psi
0},F_{\vartheta
0},F_{\zeta 0}\to 0$ as $|\varUpsilon|\to\infty$, we choose $\kappa^{-2}$ with
positive real
part,
\[ \kappa^{-2} = \sqrt{|\Pi''_o|/\overline{\gamma}_0}\ (1+ i\sgn\Pi''_o)\,, \]
noting that $\overline{\gamma}_0>0$. The two eigenfunctions which arise from the
ambiguous sign
of $\kappa$ differ only if $n$ is odd and then only in sign. Thus
\[
\kappa^{-1} = \sqrt[4]{2|\Pi''_o|/\overline{\gamma}_0}\ e^{i\pi(\sgn\Pi''_o)/8}
= \sqrt[4]{|\Pi''_o|/\overline{\gamma}_0}\bigg( \sqrt{\tfrac{\sqrt{2}+1}{2}}
+ i\sqrt{\tfrac{\sqrt{2}-1}{2}}\ \sgn\Pi''_o\bigg)\,.
\]
The $y_n$ are eigenfunctions of $\Xi$ with eigenvalue $\xi_n$,
\[
  \Xi y_n = \xi_n y_n\,,\qquad
  \xi_n = -(n+\tfrac{1}{2})\sqrt{|\Pi''_o|\overline{\gamma}_0}\
(1+i\sgn\Pi''_o) - (p_2 + i\omega_2)\,.
\]

Substitution of the ansatz \eq{2.45} into \eq{2.43} gives
\begin{align} \label{2.46}
  \mathbf{L}_n\mathbf{a} = \0\,,\qquad
  \mathbf{L}_n =
  \left( \begin{matrix}
    \xi_n & 2i\overline{\alpha}_o & 0 \\
    \Omega'_o & \xi_n & 0 \\
    \overline{W}'_o & 0 & \xi_n
  \end{matrix} \right)\,,
\end{align}
which has non-trivial solutions if $\det\mathbf{L}_n=0$. This yields $\xi_n = 0$ or
$\xi_n^2 =
2i\overline{\alpha}_o\Omega'_o$, i.e. $\xi_n =
\pm(1+i\sgn\overline{\alpha}_o\Omega'_o)\sqrt{|\overline{\alpha}_o\Omega'_o|}$, and
determines
$p_2$ and $\omega_2$. We shall ignore the solution for $\xi_n=0$, since its growth
rate has a
negative real part. The other two
solutions give for
the $n$th mode
\begin{align}
p_2 &= \mp\sqrt{|\overline{\alpha}_o\Omega'_o|} -
(n+\tfrac{1}{2})\sqrt{|\Pi''_o|\overline{\gamma}_0}
\label{2.47}\\
\omega_2 &= \mp\sqrt{|\overline{\alpha}_o\Omega'_o|}\
\sgn(\overline{\alpha}_o\Omega'_o) -
(n+\tfrac{1}{2})\sqrt{|\Pi''_o|\overline{\gamma}_0}\,\sgn\Pi''_o
\label{2.48}\\
\mathbf{a} &= [-\xi_n,\Omega'_o,\overline{W}'_o]^T\,.
\label{2.49}
\end{align}
The real and imaginary parts of the growth rate in \eq{2.47} and \eq{2.48} agree
with Gilbert
\& Ponty (2000) to order $\epsilon^2$. The vector $\mathbf{a}$ is determined up to a
constant
factor.

Equations \eq{2.40}--\eq{2.42} have solutions of the form,
\begin{align}
b_{\psi 2} &= F_{\psi 2}(\varUpsilon)e^{ik\vartheta}
+ G_{\psi 2}(\varUpsilon,\vartheta) e^{ik\vartheta}\,,\qquad
\overline{G}_{\psi 2} = 0
\label{2.50}\\
b_{\vartheta 2} &= F_{\vartheta 2}(\varUpsilon)e^{ik\vartheta} + G_{\vartheta 2}
(\varUpsilon, \vartheta) e^{ik\vartheta}\,,\qquad
\overline{G}_{\vartheta 2} = 0
\label{2.51}\\
b_{\zeta 2} &= F_{\zeta 2}(\varUpsilon)e^{ik\vartheta}
+ G_{\zeta 2}(\varUpsilon, \vartheta) e^{ik\vartheta}\,,\qquad \overline{G}_{\zeta
2} = 0\,.
\label{2.52}
\end{align}
The particular integrals $G_{\psi 2}$, $G_{\vartheta 2}$ and $G_{\zeta 2}$ can be
determined at
this order by subtracting from \eq{2.40}--\eq{2.42} their projections on
$e^{ik\vartheta}$ and
integrating with respect to $\vartheta$. The results are
\begin{equation} \label{2.53}
  \Omega_o G_{\psi 2} =
  - \frac{2\Omega'_o}{\Omega_o}\widehat{\mu}_{a,o}\varUpsilon F'_{\vartheta 0}
  + \widehat{\gamma}_0 F''_{\psi 0}
  + 2\widehat{\mu}_{a,o} F'_{\vartheta 1}
  + 2\widehat{\mu}'_{a,o}\varUpsilon F'_{\vartheta 0}
  + (2i\widehat{\alpha}_o + \widehat{\mu}_{d,o}) F_{\vartheta 0}
  + 2im\widehat{\mu}_{g,o} F_{\zeta 0}\,.
\end{equation}
\begin{equation} \label{2.54}
  \Omega_o G_{\vartheta 2} = \widehat{\gamma}_0 F''_{\vartheta 0}\,.
\end{equation}
\begin{equation} \label{2.55}
  \Omega_o G_{\zeta 2} = \widehat{\gamma}_0 F''_{\zeta 0}\,.
\end{equation}

\subsection{The asymptotic theory to $\mathcal{O}(\varepsilon^3)$ and beyond}

In this subsection we compute the solution of the $\varepsilon^3$ equations
and thus determine the magnetic field structure to $\varepsilon^1$ and 
that $p_3=\omega_3=0$. In so doing we sketch out a general method to compute
the solution at higher orders.

The $\varepsilon^3$-equations are
\begin{multline}\label{6.1}
  d_0 b_{\psi 3} + d_1 b_{\psi 2} + (d_2 - \gamma_0\d_\varUpsilon^2)b_{\psi 1}
  + (d_3 - \gamma_1\varUpsilon\d_\varUpsilon^2) b_{\psi 0} =
  2(\chi_{0,1} + \mu_{i,o})\d_\varUpsilon b_{\psi 0} \\
  + 2\mu_{a,o}\d_\varUpsilon b_{\vartheta 2} +
2\mu'_{a,o}\varUpsilon\d_\varUpsilon b_{\vartheta 1}
  + \mu''_{a,o}\varUpsilon^2\d_\varUpsilon b_{\vartheta 0}
  + (2\mu_{b,o}\d_\vartheta + 2im\mu_{c,o} + \mu_{d,o})b_{\vartheta 1} \\
  + (2\mu'_{b,o}\d_\vartheta + 2im\mu'_{c,o} + \mu'_{d,o})\varUpsilon
b_{\vartheta 0}
    +  2im\mu_{g,o} b_{\zeta 1} +  2im\mu'_{g,o}\varUpsilon b_{\zeta 0}
\end{multline}
\begin{multline}\label{6.2}
  d_0 b_{\vartheta 3} + d_1 b_{\vartheta 2} + (d_2 - \gamma_0\d_\varUpsilon^2)
b_{\vartheta 1}
  + (d_3 - \gamma_1\varUpsilon\d_\varUpsilon^2) b_{\vartheta 0} =
  2(\chi_{1,o} + \lambda_{a,o})\d_\varUpsilon b_{\vartheta 0}
  + \Omega'_o b_{\psi 1} + \varUpsilon\Omega''_o b_{\psi 0}
\end{multline}
\begin{multline}\label{6.3}
  d_0 b_{\zeta 3} + d_1 b_{\zeta 2} + (d_2 - \gamma_0\d_\varUpsilon^2) b_{\zeta
1}
  + (d_3 - \gamma_1\varUpsilon\d_\varUpsilon^2) b_{\zeta 0} =
  2(\chi_{1,o} + \rho_{b,o})\d_\varUpsilon b_{\zeta 0}\\
  + 2\rho_{a,o}\d_\varUpsilon b_{\vartheta 0}
  + \overline{W}'_o b_{\psi 1} + \varUpsilon\overline{W}_o''b_{\psi 0}\,,
\end{multline}
where we have introduced
\[ \chi_1 = \tfrac{1}{2}\nabla^2\psi + i(k\nabla\psi\cdot\nabla\vartheta +
m\nabla\psi\cdot\nabla\zeta)\,. \]
Note that we can write
$d_2 - \gamma_0\d_\varUpsilon^2 = -\Xi +
\tfrac{1}{2}\varUpsilon^2\Omega''_o(\d_\vartheta-ik) -
(\gamma_0-\overline{\gamma}_0)\d_\varUpsilon^2$.

The solutions of equations \eq{6.1}--\eq{6.3} can be written in the form,
\begin{align}
b_{\psi 3} &= F_{\psi 3}(\varUpsilon)e^{ik\vartheta} + G_{\psi
3}(\varUpsilon,\vartheta)
e^{ik\vartheta}\,,\qquad \overline{G}_{\psi 3} = 0
\label{6.4}\\
b_{\vartheta 3} &= F_{\vartheta 3}(\varUpsilon)e^{ik\vartheta} + G_{\vartheta
3}
(\varUpsilon, \vartheta) e^{ik\vartheta}\,,\qquad \overline{G}_{\vartheta 3} =
0
\label{6.5}\\
b_{\zeta 3} &= F_{\zeta 3}(\varUpsilon)e^{ik\vartheta} + G_{\zeta
3}(\varUpsilon,
\vartheta) e^{ik\vartheta}\,,\qquad \overline{G}_{\zeta 3} = 0\,.
\label{6.6}
\end{align}

Projecting equations \eq{6.1}--\eq{6.3} onto $e^{ik\vartheta}$ and using
\eq{2.13} gives
\begin{align*}
  \Xi F_{\psi 1} + 2i\overline{\alpha}_o F_{\vartheta 1}
  &= (p_3 + i\omega_3 + \tfrac{1}{6}i \Pi'''_o \varUpsilon^3 -
\overline{\gamma}_1\varUpsilon\d_\varUpsilon^2) F_{\psi 0}
  - 2(\overline{\chi}_{1,o} + \overline{\mu}_{i,o}) F'_{\psi 0}
  - 2i\overline{\alpha}'_o \varUpsilon F_{\vartheta 0}
  \\
  \Xi F_{\vartheta 1} + \Omega'_o F_{\psi 1}
  &= (p_3 + i\omega_3 + \tfrac{1}{6}i \Pi'''_o \varUpsilon^3 -
\overline{\gamma}_1\varUpsilon\d_\varUpsilon^2) F_{\vartheta 0}
  - 2(\overline{\chi}_{1,o} + \overline{\lambda}_{a,o}) F'_{\vartheta 0}
  - \varUpsilon\Omega''_o F_{\psi 0}
  \\
  \Xi F_{\zeta 1} + \overline{W}'_o F_{\psi 1}
  &= (p_3 + i\omega_3 + \tfrac{1}{6}i \Pi'''_o \varUpsilon^3 -
\overline{\gamma}_1\varUpsilon\d_\varUpsilon^2) F_{\zeta 0}
  - 2\overline{\chi}_{1,o} F'_{\zeta 0}
  - 2\overline{\rho}_{a,o} F'_{\vartheta 0} - \varUpsilon\overline{W}_o''
F_{\psi 0}\,,
\end{align*}
since $\overline{G'_{\psi 1}}=0$, $\overline{G''_{\psi 1}}=0$,
$\overline{\mu}_{a,o}=\overline{\mu}'_{a,o}=\overline{\mu}''_{a,o}=0$,
$\overline{\mu}_{d,o}=\overline{\mu}'_{d,o}=0$,
$\overline{\mu}_{g,o}=\overline{\mu}'_{g,o}=0$, $\overline{\rho}_{b,o}=0$. The
primes on the $G$ functions denote
differentiation with respect to $\varUpsilon$. Note from \eq{2.39} and
\eq{2.54},
\[
\overline{\gamma_0 G''_{\psi 1}} + 2\overline{\mu_{a,o}G'_{\vartheta 2}} =
\frac{2
F'''_{\vartheta 0}}{\Omega_o} (\overline{\gamma_0\widehat{\mu}_{a,o} +
\widehat{\gamma}_0\mu_{a,o}}) = 0\,.
\]
In vector form the projected equations are
\begin{equation}\label{6.7}
\mathbf{L} \F_1 = \{(p_3 + i\omega_3) y_n + \tfrac{1}{6}i \Pi'''_o
\varUpsilon^3y_n -
\overline{\gamma}_1\varUpsilon y''_n - 2\overline{\chi}_{1,o} y'_n \}\a -
\varUpsilon y_n\a_1 - 2y'_n\a_2\,,
\end{equation}
where $\mathbf{L}$ is defined in \eq{2.43}, $\F_1 = (F_{\psi 1},F_{\vartheta
1},F_{\zeta 1})^T$ and
\begin {equation}\label{6.8}
\a_1 = [ 2i\overline{\alpha}'_o \Omega'_o,-\Omega''_o \xi_n,-\overline{W}''_o
\xi_n]^T\,,\quad
\a_2 = [ -\overline{\mu}_{i,o} \xi_n, \overline{\lambda}_{a,o} \Omega'_o,
\overline{\rho}_{a,o} \Omega'_o ]^T\,.
\end{equation}
The primes on $y_n$ indicate derivatives with respect to $\varUpsilon$.

We next express the derivatives and terms multiplied by $\varUpsilon$ on the
right side
of \eq{6.7} in terms of parabolic cylinder functions of different orders by
using the
recurrence relations,
\begin{equation}\label{6.9}
y'_n = \tfrac{1}{2}\kappa^{-1}( n y_{n-1} - y_{n+1} )\,,\qquad
\varUpsilon y_n = \kappa ( n y_{n-1} + y_{n+1} )\,,
\end{equation}
which are derived from the parabolic cylinder function recurrence relations
 (Abramowitz \& Stegun 1972).
 Thus \eq{6.7}
becomes
\begin{equation}\label{6.10}
\mathbf{L}\F_1 = \sideset{}{'}\sum_{j=-3}^3 \mathbf{g}_{n,j}\, y_{n+j}\,,
\end{equation}
where the prime on the summation sign indicates summation over every second
index. The
vectors $\mathbf{g}_{n,j}$ are given by
\begin{align*}
\mathbf{g}_{n,0} &= (p_3 + i\omega_3) \a \\
\mathbf{g}_{n,-3} &=  \tfrac{1}{4}n(n-1)(n-2)( \tfrac{1}{3}
\overline{\gamma}_0\Pi'''_o/\Pi''_o
- \overline{\gamma}_1 )\,\kappa^{-1}\a \\
\mathbf{g}_{n,-1} &= \tfrac{1}{4}[ n^2 \overline{\gamma}_0\Pi'''_o/\Pi''_o +
n(n-2)\overline{\gamma}_1 - 4n\overline{\chi}_{1,o} ] \kappa^{-1}\a
- n \kappa\a_1 - n \kappa^{-1} \a_2 \\
\mathbf{g}_{n,1} &= \tfrac{1}{4}[ (n+1) \overline{\gamma}_0\Pi'''_o/\Pi''_o +
(n+3)\overline{\gamma}_1 + 4\overline{\chi}_{1,o} ] \kappa^{-1}\a
- \kappa\a_1 + \kappa^{-1} \a_2 \\
\mathbf{g}_{n,3} &=  \tfrac{1}{4}( \tfrac{1}{3}
\overline{\gamma}_0\Pi'''_o/\Pi''_o
- \overline{\gamma}_1 )\,\kappa^{-1}\a\,.
\end{align*}
 We assume a solution to
\eq{6.7} of the
form
\begin{equation}\label{6.11}
  \F_1 = \sideset{}{'}\sum_{j=-3}^3 \mathbf{b}_{n,j}\, y_{n+j}\,.
\end{equation}
By \eq{6.10} the coefficient vectors in \eq{6.11} are determined from the
linear equations
\begin{equation} \label{6.12}
  \mathbf{L}_{n+j} \mathbf{b}_{n,j} = \mathbf{g}_{n,j}\,, \qquad j=0,\pm 1,\pm
3\,,
\end{equation}
where $\mathbf{L}_{n+j}$ is defined in \eq{2.46}(b). When $j\neq 0$ the
determination of
$\mathbf{b}_{n,j}$ is straightforward, since $\mathbf{L}_{n+j}$ is invertible,
\begin{equation}\label{6.13}
\mathbf{L}^{-1}_{n+j} = \frac{1}{\xi_{n+j}^2-\xi_n^2}
 \begin{pmatrix}
  \xi_{n+j} & -2i\overline{\alpha}_o & 0 \\
  -\Omega'_o & \xi_{n+j} & 0 \\
  -\overline{W}'_o & 2i\overline{\alpha}_o\overline{W}'_o/\xi_{n+j} &
(\xi_{n+j}^2-\xi_n^2)/\xi_{n+j}
 \end{pmatrix}\,.
\end{equation}
We find that
\[
\mathbf{L}_{n\pm1}^{-1}\a = \mp\frac{\kappa^2}{\overline{\gamma}_0}\a\,,\qquad
\mathbf{L}_{n\pm3}^{-1}\a = \mp\frac{\kappa^2}{3\overline{\gamma}_0}\a\,,
\]
\begin{equation}\label{6.14}
\mathbf{L}_{n\pm1}^{-1}\a_1 = \frac{1}{\xi_{n\pm1}^2-\xi_n^2}
 \begin{pmatrix}
  2i\overline{\alpha}'_o\Omega'_o\xi_{n\pm1} +
2i\overline{\alpha}_o\Omega''_o\xi_n \\
  -2i\overline{\alpha}'_o(\Omega'_o)^2 - \Omega''_o\xi_n\xi_{n\pm1} \\
  -2i\overline{\alpha}'_o\Omega'_o\overline{W}'_o -
2i\overline{\alpha}_o\overline{W}'_o\Omega''_o\xi_n/\xi_{n\pm1}
  - (\xi_{n\pm1}^2-\xi_n^2)\overline{W}''_o\xi_n/\xi_{n\pm1}
 \end{pmatrix}\,,
\end{equation}
and
\begin{equation}\label{6.15}
\mathbf{L}_{n\pm1}^{-1}\a_2 = \frac{1}{\xi_{n\pm1}^2-\xi_n^2}
 \begin{pmatrix}
  -\overline{\mu}_{i,o}\xi_n\xi_{n\pm1}
  - 2i\overline{\alpha}_o\overline{\lambda}_{a,o}\Omega'_o \\
  \Omega'_o\overline{\mu}_{i,o}\xi_n
  + \overline{\lambda}_{a,o}\Omega'_o\xi_{n\pm1} \\
  \overline{W}'_o\overline{\mu}_{i,o}\xi_n
  +
2i\overline{\alpha}_o\overline{W}'_o\Omega'_o\overline{\lambda}_{a,o}/\xi_{n\pm1}
  + (\xi_{n\pm1}^2-\xi_n^2)\overline{\rho}_{a,o}\Omega'_o/\xi_{n\pm1}
 \end{pmatrix}\,.
\end{equation}
Hence the solution vectors in \eq{6.11} determined from \eq{6.12} for $j\neq0$
are
\begin{align}
\mathbf{b}_{n,-3} &=  \tfrac{1}{12}n(n-1)(n-2)( \tfrac{1}{3}
\overline{\gamma}_0\Pi'''_o/\Pi''_o
- \overline{\gamma}_1 )\frac{\kappa}{\overline{\gamma}_0}\,\a
\label{6.16}\\
\mathbf{b}_{n,-1} &= \tfrac{1}{4}[ n^2 \overline{\gamma}_0\Pi'''_o/\Pi''_o
+ n(n-2)\overline{\gamma}_1 - 4n\overline{\chi}_{1,o} ]
\frac{\kappa}{\overline{\gamma}_0}\a
- n\mathbf{L}_{n-1}^{-1}(\kappa\a_1 + \kappa^{-1} \a_2)
\label{6.17}\\
\mathbf{b}_{n,1} &= -\tfrac{1}{4}[ (n+1) \overline{\gamma}_0\Pi'''_o/\Pi''_o
+ (n+3)\overline{\gamma}_1 + 4\overline{\chi}_{1,o} ]
\frac{\kappa}{\overline{\gamma}_0}\a
- \mathbf{L}_{n+1}^{-1}(\kappa\a_1 - \kappa^{-1} \a_2)
\label{6.18}\\
\mathbf{b}_{n,3} &=  -\tfrac{1}{12}( \tfrac{1}{3}
\overline{\gamma}_0\Pi'''_o/\Pi''_o
- \overline{\gamma}_1 )\frac{\kappa}{\overline{\gamma}_0}\a\,.
\label{6.19}
\end{align}

In the $j=0$ case the matrix $\mathbf{L}_n$ is singular. It satisfies
$\c^T\mathbf{L}_n=\0$, where
\begin{equation}\label{6.19.1}
    \c=[-\Omega'_o\,,\xi_n\,,0]^T/2\Omega'_o \xi_n
\end{equation}
and $\c^T\a=1$. Thus the $j=0$ equation in \eq{6.11} furnishes us with the
solvability
condition $\c^T\mathbf{g}_{n,0} = 0$, which gives us
\begin{equation}
 p_3 + i\omega_3 = 0,
\end{equation}
and which ensures
that
$\mathbf{g}_{n,0}=\mathbf{0}$ and $\mathbf{b}_{n,0}$ is a constant
multiple of $\a$.
Thus the term $\mathbf{b}_{n,0}\, y_n$ can be absorbed into the order
$\varepsilon^0$
solution. Without loss of generality we can set $\mathbf{b}_{n,0}=\0$.
The particular $G$ solutions can consequently be computed, but we omit those
details in the interest of space.

 Higher order equations can be reduced to the
form above and solved similarly, though some fortitude is required to weather
the algebraic maelstrom that ensues. Generally, at some order
$\varepsilon^{N+2}$, the projected inhomogeneous equations can be manipulated
into
$$ \mathbf{L}\,\mathbf{F}_N =
\sideset{}{'}\sum_{j=-3N}^{3N}\,\mathbf{h}_{n,j}\,y_{n+j},$$
which admits a solution of the form 
$$ \mathbf{F}_N =
\sideset{}{'}\sum_{j=-3N}^{3N}\,\mathbf{b}_{n,j}\,y_{n+j},$$
with the $\mathbf{b}_{n,j}$ coefficients determined by linear equations, and
the growth rate and frequency $p_{N+2}$ and $\omega_{N+2}$ determined from the
singular $j=0$ case.

\section{Numerical evaluation of the asymptotic expressions}

In general the asymptotic approximations \eq{fuck1}-\eq{2.58} to the
growth rates
and the eigenfunctions must be evaluated numerically. In this section we describe
the method of
computation.

The formula for the growth rate requires the evaluation of $ \Omega$, $\Omega'$,
$\Omega''$,
$\overline{W}$, $\overline{W}'$ , $\overline{W}''$, $\overline{\mu}_b$,
$\overline{\mu}_c$,
$\overline{\beta}_k$, $\overline{\beta}_m$, $\overline{\beta}_{mk}$ and
$\overline{\gamma_0}$
on the resonant streamline $\Psi=\Psi_o$. Each of these quantities may be evaluated
by line
integrals along the streamline. Moreover a number of their constituent parts (such
as $\nabla
\vartheta$ and $\nabla \zeta$) may also be determined by line integrals. The
integrals are
 evaluated numerically using the
compound trapezoidal rule. The eigenfunctions require the evaluation of these
quantities,
except the $\beta$'s, on a $(\vartheta,\Psi)$-grid, which must subsequently be
interpolated
onto the $(s,z)$ coordinate system. A simple linear interpolation was sufficient.

We obtain $\Omega$ by integrating \eq{2.3} and $\overline{W}$ by averaging. Their
$\Psi$
derivatives may be procured as line integrals using the following technique. The
average of a
function $F(\Psi,\vartheta)$ over the curve $ C_\Psi$ given by constant $\Psi$ can be
expressed, using \eq{2.3}, as
\begin{align} \label{3.1}
\overline{F}(\Psi) =  \frac{\Omega}{2\pi} \oint_{ C_\Psi}\frac{F\,\vm}{q^2}\cdot d\r
&= \frac{\Omega}{2\pi} \oint_{\d S_\Psi} \frac{F\,\vm}{q^2}\cdot d\r
+ \frac{\Omega}{2\pi} \oint_{C_0} \frac{F\,\vm}{q^2}\cdot d\r
\nonumber\\
&= \frac{\Omega}{2\pi} \oint_{S_\Psi} \nabla\times \left(
\frac{F}{q^2}\vm \right) \cdot \1_\phi \,r\,dr\,d\theta + \Omega K\,,
\end{align}
where $S_\Psi$ is the annular region in the meridional plane bounded by $ C_\Psi$
and a smaller
fixed $\Psi$-curve $C_0$, which encloses the stagnation point. The quantity $K$ is
independent
of $\Psi$. Now by \eq{2.7}, $r dr\,d\theta  = (\Omega\, r\sin\theta)^{-1}
d\Psi\,d\vartheta$,
and thus
\begin{equation} \label{3.2}
\frac{\overline{F}}{\Omega} = \frac{1}{2\pi} \oint_{S_\Psi}
\nabla\times\left( \frac{F}{q^2}\vm \right) \cdot \frac{\1_\phi}{\Omega
r\sin\theta}\, d\vartheta \,d\Psi + K\,.
\end{equation}
Differentiation gives
\begin{equation}\label{3.3}
\frac{d}{d\Psi}\left(\frac{\overline{F}}{\Omega}\right) = \frac{1}{2\pi}
\oint_{C_\Psi} \nabla\times\left( \frac{F}{q^2} \vm \right) \cdot
\frac{\1_\phi}{\Omega r\sin\theta}\,d\vartheta
= \frac{\overline{F_1}}{\Omega}\,,
\end{equation}
where $F_1=\nabla\times(F\vm/q^2)\cdot\1_\phi/r\sin\theta$. Iteration yields
\begin{equation}\label{3.4}
  d^n(\overline{F}/\Omega)/d\Psi^n = \overline{F_n}/\Omega\,,
\end{equation}
where $F_n$ is defined inductively for integer $n>1$ by
\begin{equation}\label{3.5}
  F_n = \nabla\times(F_{n-1}\vm/q^2)\cdot\1_\phi/r\sin\theta
  = -\nabla\cdot\bigg(\frac{F_{n-1}\nabla\Psi}{(\nabla\Psi)^2}\bigg)\,.
\end{equation}

Setting $F=1$ gives integral expressions for $\Omega'$, $\Omega''$, etc. For
azimuthal flows of
the form $W(\vartheta,\Psi)$, $F=W$ gives expressions for $\overline{W}$,
$\overline{W}'$,
$\overline{W}''$, etc. The integrands rapidly become very complicated with $n$. The
motivation
for persisting with these complicated expressions is that the numerical integration
of smooth
periodic functions over a period using the compound trapezoidal rule is spectrally
accurate.

It soon becomes apparent, however,
 that a number of the integrands are singular in spherical
polar
coordinates. If $\V_m = V_r\,\1_r + V_\theta\,\1_\theta$, we can write $d\vartheta =
\Omega r
d\theta/V_\theta = \Omega dr/V_r$. The spherical polar components of $\V_m$ vanish
at two
points on a $\V_m$-streamline, which can cause singular integrals, specifically
those which
evaluate $\nabla\vartheta$. To avoid this problem we transform to the toroidal
coordinate
system $(R,\Theta)$ shown in Figure~9. The point $P_0(r_0,\theta_0)$ is the stagnation
point of
the meridional flow.

\begin{figure}
\begin{center}
\begin{pspicture}(5,5)
    \psset{linewidth=.8pt}
    \psaxes[labels=none,ticks=none]{->}(0,0)(5,5)
    \uput{0.2}[d](4.5,0){$s$}
    \uput{0.2}[l](0,4.5){$z$}
    \psarc(0,0){4}{0}{90}
    \psline(0,0)(2,0.5)
    \psline(2,0.5)(3,2)
    \psline(0,0)(3,2)
    \psline[linewidth=0.5pt](2,0.5)(2.75,0.5)
    \qdisk(2,0.5){1pt}
    \uput{0.15}[-45](2,0.5){$P_0$}
    \qdisk(3,2){1pt}
    \uput{0.1}[135](3,2){$P$}
    \psarc[linewidth=0.5pt]{->}(2,0.5){0.5}{0}{56.310}
    \uput{0.6}[22](2,0.5){$\Theta$}
    \uput{1}[45](2,0.5){$R$}
    \psarcn[linewidth=0.5pt]{->}(0,0){0.5}{90}{14.036}
    \uput{0.6}[52](0,0){$\theta_0$}
    \uput{1.1}[7](0,0){$r_0$}
    \psarcn[linewidth=0.5pt]{->}(0,0){1.2}{90}{33.690}
    \uput{1.3}[61](0,0){$\theta$}
    \uput{2}[40](0,0){$r$}
\end{pspicture}
\caption{Toroidal coordinates $(R,\Theta)$.}
\end{center}
\end{figure}
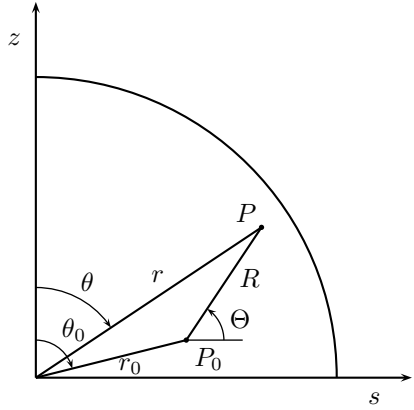

Basic trigonometry gives a number of relationships between $(r,\theta)$ and
$(R,\Theta)$, the
most important of which are
\begin{align}
r &= \sqrt{r_0^2 + R^2 + 2r_0R\sin(\theta_0 + \Theta)}
\label{3.7}\\
\theta &= \theta_0 + \tan^{-1}\left| \frac{R\cos(\theta_0 + \Theta)}{r_0 +
R\sin(\theta_0 +
\Theta)} \right|\,, \label{3.8}
\end{align}
The $(R,\Theta,\phi)$ system is orthogonal but left-handed. For the flows we
examine, we
take $(r_0,\theta_0)$ to be the stagnation point at the centre of the concentric
$\Psi$ curves.
The radius $r_0$ can be evaluated using the Newton-Raphson method. We distinguish
between
different $\Psi$ curves by their largest $s$-intercept $r_s$. Each $\Psi$ curve is thus
described by an equation $\Psi(r,\theta) = \Psi(r_s,\pi/2)$. To determine the
quadrature nodes
we divide $[0,2\pi]$, corresponding to one period of $\Theta$, into $K$ equal
sub-intervals
using the nodes $ \Theta_i = 2\pi i/K$, $i=0,1,\dots,K$. To calculate the
corresponding values
of $R_i$ we use the Newton-Raphson method to solve
\[ \Psi(r(R_i,\Theta_i),\theta(R_i,\Theta_i)) - \Psi(r_s,\pi/2) = 0 \]
for $R_i$ given $\Theta_i$.

The integrands $\mu_b$, $\mu_c$, $\beta_k$, $\beta_m$, $\beta_{mk}$ and $\gamma_0$ are
expressed in terms of: $\Omega$, $\overline{W}$; the $R$ and $\Theta$ derivatives of
$\Psi$,
$\vartheta$, $V_\Theta$, $V_R$; and the $\Psi$ and $\vartheta$ derivatives of $Z$;
where $\V_m
= V_R\,\1_R + V_\Theta\,\1_\Theta$. The derivatives of $\Psi$, and $V_R$, $V_\Theta$
can be
found analytically. The quantities $\d_R \vartheta$, $\d_\Theta \vartheta$ and
$\d_\Psi Z$ must
be determined numerically. The angle $\vartheta$ is given by
\begin{equation}  \label{3.9}
\vartheta = \int_0^\vartheta d\vartheta^* = -\Omega \int_0^\Theta R^*
\frac{(r_0\sin\theta_0+R^*\cos\Theta^*)}{\d_{R^*}\Psi^*} d\Theta^*\,,
\end{equation}
where an asterisk denotes evaluation upon the $\Psi=\Psi^*$ curve. The two sets of
variables
$(R,\Theta)$ and $(R^*,\Theta^*)$ should not be confused: the asterisked pair are
dependent on
each other while the other pair are independent. With this in mind we differentiate the
integral in \eq{3.9}. Using Leibniz's theorem we obtain
\begin{equation} \label{3.10}
\d_R\bigg(\frac{\vartheta}{\Omega}\bigg) = -\int_0^\Theta \d_{R^*}
\bigg( \frac{R^* (r_0\sin\theta_0 + R^*\cos\Theta^*)}{\d_{R^*}\Psi^*} \bigg)
\frac{\d_R\Psi}{\d_{R^*} \Psi^*} d\Theta^*\,,
\end{equation}
and
\begin{equation} \label{3.11}
\d_\Theta\bigg(\frac{\vartheta}{\Omega}\bigg) = -\int_0^\Theta \d_{R^*}
\bigg( \frac{R^* (r_0\sin\theta_0 + R^*\cos\Theta^*)}{\d_{R^*}\Psi^*} \bigg)
\frac{\d_\Theta\Psi}{\d_{R^*}\Psi^*} d\Theta^*
- \frac{R (r_0\sin\theta_0 + R\cos\Theta)}{\d_R\Psi}\,.
\end{equation}

Lastly, $\d_\Psi Z$ and $\d_\vartheta Z$ are required for $\nabla\zeta$. Only the
former issues
a challenge. From the definition of $\widetilde{W}$,
\begin{equation} \label{3.12}
\d_\Psi(\Omega Z) = \int_0^\vartheta \d_\Psi \widetilde{W}^* d\vartheta^*
= \int_0^\vartheta \d_\Psi W^* d\vartheta^* - \overline{W}'\vartheta\,.
\end{equation}
Thus from \eq{2.4},
\begin{equation} \label{3.13}
\d_\Psi Z = \frac{1}{\Omega} \int_0^\vartheta (\d_\Psi W^* - \frac{\Omega'}{\Omega}
W^*)d\vartheta^*
+ \frac{\vartheta}{\Omega} \left(\frac{\Omega'}{\Omega} \overline{W} -
\overline{W}'\right)\,.
\end{equation}
This expression is evaluated by converting the $\vartheta$ integral to an integral over
$\Theta$ and using the formula
\begin{equation} \label{3.14}
\d_\Psi W = J^{-1} (\d_\Theta\vartheta\,\d_R W - \d_R\vartheta\,\d_\Theta W)\,,\qquad
J = \frac{\d(\Psi,\vartheta)} {\d(R,\Theta)}\,.
\end{equation}
Note that as these integrals are not over closed curves, the trapezoidal rule does not
yield exponential accuracy. The $\mu$'s and $\beta$'s also require further
averaging so their convergence is not as fast (as shown in Table~3).

\begin{table}
\begin{center}
\begin{footnotesize}
\begin{tabular}{c|rrrr}
\hline
$K$                & 100        & 200        & 400        & 800 \\
\hline
$\overline{\mu}_b$ & 4.63875    & 4.63936    &  4.63951   & 4.63955 \\
$\overline{\mu}_c$ & $-0.20886$ & $-0.20782$ & $-0.20756$ & $-0.20750$ \\
$\beta_k$          & 8.961      & 8.964      & 8.965      & 8.965 \\
$\beta_m$ ($\v_1$) & 4.58627    & 4.58627    & 4.58627    & 4.58627 \\
$\beta_m$ ($\v_2$) & 4.6719     & 4.6718     & 4.6718     & 4.6718 \\
$\beta_{mk}$       & 0.2256     & 0.2278     & 0.2284     & 0.2285 \\
\hline
\end{tabular}
\end{footnotesize}
\end{center}
\caption{The quantities required by the asymptotic theory which converge most slowly
with $K$.
Here $r_s=0.93$ and $K$ is the number of subintervals approximating $\Psi_o$.}
\end{table}

The asymptotic estimates were computed in MATLAB. The convergence of quantities
required by the
asymptotic theory is shown in Table~3 for different numbers $K$ of numerical
integration nodes
along the chosen streamline, $\Psi_o$. Those quantities that issue from a single
integration
around the closed streamline converge very rapidly, typically for $K=30$. These include
$\Omega_o$ and its $\Psi$ derivatives, $\overline{W}$ and its $\Psi$ derivatives, and
$\overline{\gamma}_0$. Thus the quantities $\Omega_o=5.3919$, $\Omega'_o=7.2807$,
$\Omega''_o=-16.662$, $\overline{W}_o=0.93043$, $\overline{W}'_o=1.4927$,
$\overline{W}'''_o=-2.7786$ are accurate to the figures shown here for $K=100$.
However,
quantities which are evaluated by line integrals with variable limits converge more
slowly, and
settle down only for $K=800$.

\end{document}